\documentclass[journal, final]{IEEEtran}
\normalsize
\ifCLASSINFOpdf
\else
\fi

\usepackage{comment}
\usepackage{enumitem}
\usepackage{float}

\usepackage{amsthm}
\usepackage{amsmath}
\usepackage{amssymb}
\usepackage{graphicx}
\usepackage{esint}
\usepackage{amsfonts}
\usepackage{cite}
\usepackage{balance}
\usepackage{caption}
\usepackage{subcaption}
\usepackage{epstopdf}
\usepackage{color}
\usepackage{tikz}
\usepackage{pgfplots} 
\usetikzlibrary{patterns}
\makeatletter

\theoremstyle{plain}
\newtheorem{thm}{\protect\theoremname}

\makeatother

\providecommand{\theoremname}{Theorem}

\theoremstyle{plain}
\newtheorem{prop}{\protect\propname}

\makeatother

\providecommand{\propname}{Proposition}

\theoremstyle{plain}

\makeatother

\providecommand{\lemmaname}{Lemma}

\theoremstyle{plain}

\makeatother

\providecommand{\remarkname}{Remark}

\begin{document}
\title{Holographic RIS Empowered THz Communications with Hardware Imperfections under Adverse Weather Conditions}

\author{ Alexandros-Apostolos~A.~Boulogeorgos,~\IEEEmembership{Senior Member,~IEEE,} Stylianos~Trevlakis,~\IEEEmembership{Member,~IEEE,} and Theodoros A. Tsiftsis,~\IEEEmembership{Senior Member,~IEEE}

\thanks{A.-A. A. Boulogeorgos is with the Department of Electrical and Computer Engineering, University of Western Macedonia, 50100 Kozani, Greece (e-mail:
al.boulogeorgos@ieee.org).
}
\thanks{S. Trevlakis is with the Research \& Development Department, InnoCube P.C., Thessaloniki 55535, Greece. (e-mail: trevlakis@innocube.org).
}
\thanks{Theodoros A. Tsiftsis is with the Department of Informatics \& Telecommunications, University of Thessaly, Lamia 35100, Greece (e-mail:tsiftsis@uth.gr).
}
}
\maketitle

\begin{abstract}
This paper focuses on providing a theoretical framework for the evaluation of the performance of holographic reconfigurable intelligent surface (HRIS) empowered terahertz (THz) wireless systems under fog conditions. In more detail, we present a comprehensive methodology for evaluating the geometric losses of the end-to-end channel. Moreover, the stochastic nature of the end-to-end channel is characterized by novel closed-form probability density and cumulative distribution functions. Building upon them, the outage probability and the throughput of the system are extracted in closed form. These formulas account for the impact of transceiver hardware imperfections and are expected to become useful tools for the design of HRIS empowered THz wireless systems due to its remarkable engineering insights.       
\end{abstract}


\section{Introduction}\label{S:Intro}
\subsection{State of the Art \& Motivation}
Bringing the fiber quality of experience into the wireless world through the development of wireless fiber extenders capable of supporting data rates in the Tb/s region is a fundamental promise of terahertz (THz) technologies~\cite{Boulogeorgos2018}. THz wireless fiber extenders aim to establish low-cost medium to long range high-bandwidth links for mid- and back-hauling without demanding the time-consuming processes of optical fiber installation~\cite{WP:Wireless_Thz_system_architecture_for_networks_beyond_5G,Boulogeorgos2021a}. As a consequence, from a business point of view, flexible almost-instance high-data-rate connectivity is expected to bring important economic benefits, while allowing internet democratization~\cite{8782882,8808165}.

Nevertheless, THz wireless communication systems are susceptible to a number of drawbacks. First of all, they suffer from high pathloss due to the high-transmission frequencies as well as molecular resonances that lead to molecular adsorptions~\cite{8904434}. Motivated by this, a great amount of research effort was put on modeling the deterministic path loss of THz wireless links~\cite{5995306,7086348,7955066,8568124}. In more detail, in~\cite{5995306}, the authors employed the radiation theory in order to extract the channel characteristics of nano-scale THz wireless links. In~\cite{7086348}, an electromagnetic analysis was conducted for the propagation characterization of in-body THz links, whilst in~\cite{7955066}, the authors presented a path loss model for in-body nano-scale THz wireless systems. A simplified path loss model for THz systems operating in the $275-400\,\rm{GHz}$ region was presented in~\cite{8568124}. 

Another important drawback of THz wireless systems is the signal fluctuation due to atmospheric conditions. The source of this phenomenon lies in the variation of the reflection index, as a result of  inhomogeneities in atmospheric pressure as well as temperature and water molecular density across the propagation path~\cite{1638639}.  Inspired by this, the authors of~\cite{Ma2015} used a weather emulating chamber in order to quantify in laboratory conditions the impact of rain in THz and infrared links. In~\cite{9787400}, the authors quantified the joint impact of rain and misalignment in THz wireless systems in terms of outage probability. The detrimental impact of turbulence in THz wireless systems was discussed in~\cite{8751955}. The joint impact of snow and misalignment in THz wireless systems was evaluated in terms of bit error rate and channel capacity in~\cite{Taherkhani:20}. The authors of~\cite{Su:12} quantified the impact of fog in THz wireless systems in terms of bit error rate in laboratory environment. Finally, in~\cite{6971163}, the authors experimentally demonstrated the distortion to the THz signal that was caused due to fog. 

To counterbalance the significant path loss as well as the signal fluctuation due to the atmospheric conditions and establish long-range connections, high-directional antennas, such as Cassegrain antennas, are employed at both the transmitter (TX) and the receiver (RX) ends, establishing high-directional links~\cite{Boulogeorgos2019}. However, these links are sensitive to blockage, due to the high penetration loss of the THz transmission. Therefore, the need to create alternative paths between the TX and RX~arises.  On another front, with the explosive growth of data and bandwidth requirements in beyond fifth generation (5G) and sixth generation (6G) wireless communication networks, there is a unprecedented demand for improving key performance indicators (KPIs) such as energy efficiency, spectral efficiency, coverage and serviceability at very high levels~\cite{3gppR18, Mathaiou_COMMAG}. 

To this end, reconfigurable intelligent surfaces (RIS) are emerging as the revolutionary technology based on programmable metamaterial-based surfaces and seem to be the fundamental component of 6G communication infrastructures in improving the above mentioned KPIs \cite{Renzo2020, j:whitepaper2023}. In particular, conventional RIS plates comprises an array of nearly passive reflecting meta-elements, which are able to control impinging signals through a field programmable gate array (FPGA) microcontroller. As a result, RIS adapts both the amplitude and phase of incident signals to the specific wireless environment conditions in order to either increase the coverage or cancel interference, and highlight its effectiveness in various applications such as drone-based communications, integrated sensing and communications, full-duplex communications, etc. \cite{9530717, 9903378, 9919748}. 

Above and beyond all other considerations, due to the adaptability of RIS in creating alternative wireless paths in highly penetrated path loss environments i.e., THz communication systems, RIS appears to be a reliable technology to to combat communication blockage~\cite{Basar2019, 9433568}. Specifically, the authors in \cite{Basar2019} presented for the first time key performance results (path loss and error rate) of RIS-enabled wireless communication systems, and conducted theoretical comparisons with well-established technologies such as such as relaying, multiple-input multiple-output (MIMO) 
beamforming, and backscatter communications. In \cite{9433568}, the path loss of RIS-assisted wireless  
systems was analytically investigated based on the vector generalization of Green’s theorem. In~\cite{9881509}, an improved path loss model for RIS-enabled wireless communications at mmWave bands was proposed. In particular, practical electromagnetic phenomena (e.g., gain patterns, received power, phase errors, and specular reflection loss per meta-element), were investigated. A similar approach is followed in \cite{9837936} where an angle-dependent  path loss model for RISs was proposed based on the radiation patterns of all involved parts of the communication system. However, the proposed prototype in \cite{9881509, 9837936} are based on PIN diodes, which is impractical for sub-THz bands \cite{10159567}. Furthermore, RIS is highly affected by the frequency range of the incident signal to the RIS meta-elements. Therefore, RIS's reflection and phase adjustment capability employing either varactors or PIN diodes is limited over RF frequency range and, specifically, between 100 MHz - 10 GHz. Therefore, for much higher frequencies at the sub-THz bands, i.e., 0.1-0.3 THz, micro-electromechanical system (MEMS), mechanical approach, liquid crystal, and microﬂuidics appear to be the most suitable tuning RIS technologies \cite{9690474}. It is worth mentioning that the latter two tuning technologies cover a wide range of THz bandwidths compared to the two first mentioned tuning ones   \cite{7109827, 10159567}. The operation of RIS at sub-THz is also feasible with tuning technologies involving complementary metal-oxide-semiconductor (CMOS) transistors, Schottky diodes, 
or high electron mobility transistors (HEMTs) \cite{nature, 10159567}. Very recently, the authors in \cite{Nature_Light}, have demonstrated for the first time a THz point-to-point RIS-aided transmission based on real-time beam tracking. In this experiment, the THz signals  near  $0.34$ THz are controlled via a RIS plate consisted of GaN HEMTs

Despite the fact that RIS is potentially considered as the key enabling technology to drive the evolution of beyond 5G and 6G era, it is true that confronts critical challenges. For example, its nearly passive nature and the absence of any active components makes conventional RIS to perform channel estimation and beam tracking as one of the most challenging issues. Consequently, in real life setups, conventional RISs constraint the bandwidth used for transmission and the achievable data rate. 
To address the above issues in conventional RIS, the holographic RIS  (HRIS) (a.k.a. as reconfigurable holographic surface), has been emerged recently as an alternate technology \cite{Huang2020, Deng2021, 9690474}. HRIS is considered as a limited surface area that integrates a virtually inﬁnite number of tiny metamaterial-based radiation elements in order to form a spatially continuous transceiver aperture that can achieve the holographic beamforming. HRIS is able to support channel estimation and act like continuous surface for higher frequencies up to the THz bandwidth compared to conventional RIS \cite{9690474}. 

In one of the most representative works on HRIS \cite{9374451}, the authors  proposed the concept of HRIS  and studied its application to massive MIMO in THz frequency range. More, in this pioneer work, the authors have addressed the transmission design of nearly-passive HRISs with 
spatially continuous apertures by designing the beam pattern of a RIS with discrete meta-elements and generalize their beamforming framework to a surface with closely spaced (i.e. continuous) meta-elements yielding the HRIS design. In \cite{9903514}, the authors studied the problem of beam pattern design of RIS-assisted THz communication systems with two-dimensional ﬁnite impulse response ﬁlter design, and the above method proved to facilitate applications such as wireless positioning and over-the-air computation.
In \cite{10087279}, the joint phase and delay wideband precoding was investigated to tackle the severe problem of array gain loss of RIS-aided THz communications due to beam split at RIS. Additionally, the authors in \cite{10001180} have investigated the performance of HRIS non-orthogonal multiple access (NOMA) networks, whereas the same authors in \cite{10177184} have very recently evaluated the outage of HRISs-enabled THz NOMA systems in the presence of misalignment errors and for either perfect or imperfect successive interference cancellation. 

Motivated by the fact that HRIS supports the same functionalities as conventional RIS, but most importantly due to its holographic pattern is an attractive candidate for RIS-aided wireless communication systems working at the sub-THz band, in this paper we study  the performance of HRIS-empowered THz wireless systems with hardware imperfections at both ends and under foggy weather conditions.

\subsection {Our Contributions}
In contrast that most of the published works study the application of conventional RIS in either sub-6GHz bands or mmwave, in this paper, the communication theoretical framework of HRIS in sub-THz bands over severe weather conditions (fog) is analytically studied under the presence of transceiver hardware imperfections. The contributions of our work can be summarized as follows:
\begin{itemize}
\item The end-to-end geometric losses of the proposed system are calculated by considering the  following channel coefficients: free-space geometric channel gain, molecular absorption gain, and fog gain. Our analysis shows that the geometric losses are quite affected by the following parameters: the relative position of HRIS in respect to the TX, HRIS size and its orientation, TX/RX antenna gains, the signal frequency, as well as the weather conditions.

\item The statistics of the end-to-end channel coefficients of HRIS-enabled THz systems is analytically evaluated in closed form. Based on the latter derivations, new closed-form expressions for the outage probability and data throughput are extracted when hardware imperfections take place at both ends of the considered system in foggy weather conditions.
 \item New engineering insights are extracted such as how the geometric loss is affected by various frequencies in the sub-THZ band, and what is the optimal placement of HRIS between transmitter and receiver. Additionally, our paper provides a comprehensive study on the impact of the placement and orientation of HRIS to both outage and throughput performance metrics under fixed values of frequency range, spectral efficiency, and transceiver hardware imperfection parameters.
\end{itemize}

\subsection{Organization of the paper}	
This paper is structured as follows. In the sequel, a few basic mathematical symbols used in this paper are given. The system and signal model studied in this paper are introduced in Section II. In Section III, the geometric losses and the corresponding statistical analysis of the end-to-end channels are analytically presented, while both the outage probability and throughput are obtained in closed form. In Section IV, the analytical expressions are corroborated and cross-compared with Monte Carlo simulation results. In the same Section, some useful engineering insights are revealed and discussed in much detail. Finally, some concluding remarks are given in Section V.

\subsection{Notations}
In this paper, $\sqrt{x}$ returns the square root of $x$. The cosine and exponential functions are denoted by $\cos\left(\cdot\right)$ and $\exp\left(\cdot\right)$, respectively. The natural logarithm is represented as $\ln(\cdot)$, while $\log_{a}(\cdot)$ is the logarithm to base $a$. The operator $\min\left(a, b\right)$ returns the minimum between $a$ and $b$. The gamma and the upper incomplete gamma functions are respectively represented as $\Gamma(\cdot)$~\cite[eq. (8.310)]{B:Gra_Ryz_Book} and $\Gamma\left(\cdot,\cdot\right)$~\cite[eq. (8.350/2)]{B:Gra_Ryz_Book}. The Kummer confluent hypergeometric function is denoted by $\,_1F_1\left(\cdot,\cdot,\cdot\right)$~\cite[eq. (9.14/1)]{B:Gra_Ryz_Book}. Moreover, $H_A\left(a,b;c;x_1, x_2\right)$ stands for the Lauricella hypergeometric function~\cite{CARLSON1963452}. Finally, the Pochhammer operator is denoted by $(n)_m$. 

\section{System and signal model}\label{sec:SSM}
\begin{figure}
    \centering
    \includegraphics[width=1\linewidth]{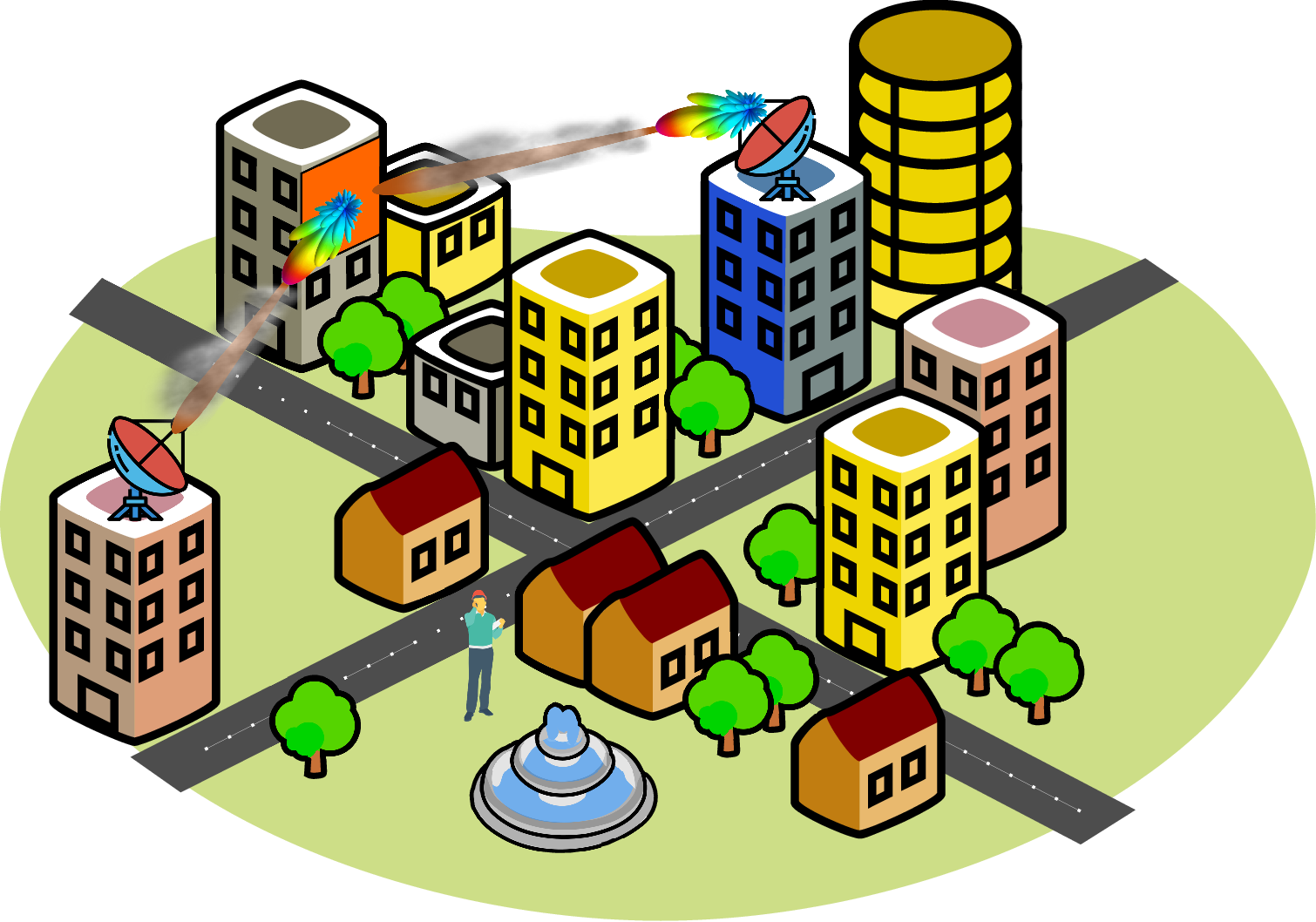}
    \caption{Typical model of an HRIS empowered THz communication system.}
    \label{Fig:System_model}
\end{figure}
As demonstrated in Fig.~\ref{Fig:System_model}, we consider a long-range THz wireless fiber extender that is established through a HRIS. By assuming that the transmitter (TX)-HRIS and HRIS-receiver (RX) links experience fog, the baseband equivalent received signal can be expressed~as
\begin{align}
    r = h_g\,h_1 \, h_2 \, (s + n_t) + n_r + n, 
    \label{Eq:r}
\end{align}
where $h_g$ stands for the end-to-end geometric gain channel coefficient, while $h_1$ and $h_2$ are independent random processes that model the impact of fog in the TX-HRIS and HRIS-RX links, respectively. The transmitted signal is denoted by $s$, whereas $n_t$ and $n_r$ stands for the distortion noises due to hardware imperfections at the TX and RX, respectively. The additive white Gaussian noise is modeled by the zero-mean complex Gaussian random process $n$ of variance $\sigma_n^2$. 

The end-to-end geometric gain channel coefficient can be analyzed~as
\begin{align}
    h_g = h_{g,f}\,h_{g,m}\,h_{f},
    \label{Eq:hg}
\end{align}
where $h_{g,f}$ is the free space geometric gain channel coefficient that can be expressed~as~\cite[eq. (9)]{8936989}
\begin{align}
    h_{g,f} = \frac{c\,\sqrt{G_t\,G_r}\,l_h\,l_v}{4\pi\,f\,d_1\,d_2}\cos\left(\psi\right),
    \label{Eq:h_gf}
\end{align}
with $c$ being the speed of light, $G_t$ and $G_r$ standing for the transmission and reception antenna gains, respectively, while $l_h$ and $l_v$ representing the length and width of the HRIS. Also, $f$ is the transmission frequency, whereas $d_1$ and $d_2$ are the TX-HRIS and HRIS-RX distances, respectively. Finally, $\psi$ denotes the beam incident angle at the HRIS.  

Additionally, $h_{g,m}$ is the molecular absorption gain coefficient that can be obtained~as
\begin{align}
    h_{g,m} = \exp\left(\frac{1}{2}\,\kappa_m\left(f, T, \phi, p\right)\,\left(d_1+d_2\right)\right),
    \label{Eq:h_gm}
\end{align}
where $\kappa_m$ is the molecular absorption coefficient, $T$ is the atmospheric temperature, $\phi$ stands for the relative humidity, and $p$ represents the atmospheric pressure. 

Based on~\cite{5995306}, the molecular absorption coefficient can be written~as
\begin{align}
    \kappa_m\left(f, T, \phi, p\right)=\sum_{k,l} \kappa^{g}_{k,l}\left(f, T, \phi, p\right),
\end{align}
where $ \kappa^{g}_{k,l}\left(f, T, \phi, p\right)$ represents the molecular absorption coefficient for the isotoplogue $k$ of the $l-$th gas and can be expressed~as
\begin{align}
    \kappa^{g}_{k,l}\left(f, T, \phi, p\right) = \frac{p}{p_o} \, \frac{T_o}{T} Q^{g}_{k,l}\,\sigma_{k,l}^{g}\left(f, T, \phi, p\right),
\end{align}
with $p_o$ and $T_o$ denoting the standard pressure and temperature respectively, while $Q^{g}_{k,l}$ and $\sigma_{k,l}^{g}\left(f, T, \phi, p\right)$ respectively being the molecular volumetric density and the absorption cross section for the isotopologue $k$ of the $l-$th gas. The molecular volumetric density can be evaluated~as
\begin{align}
    Q^{g}_{k,l} = \frac{p}{R\,T}\,q_{k,l}\,N_{A},
\end{align}
where $R$ is the gas constant, $N_A$ stands for the Avogadro constant, and $q_{k,l}$ represents the mixing ratio for the isotopologue $k$ of the $l-$th gas. Note that in the high resolution transmission (HITRAN) database, each isologue contribution is scaled based on its natural abundance in the medium. As a consequence, instead of $q_{k,l}$, the mixing ratio of the specific gas is employed. 

The absorption cross section for the isotopologue $k$ of the $l-$th gas can be calculated~as 
\begin{align}
    \sigma_{k,l}^{g}\left(f, T, \phi, p\right) = S_{k,l}\,G_{k,l}\left(f, T, \phi, p\right),
\end{align}
where $S_{k,l}$ is the absorption strength of a specific type of molecule and can be directly extracted from the HITRAN database, while $G_{k,l}\left(f, T, \phi, p\right)$ is the spectral line shape for the isotopologue $k$ of the $l-$th gas that can be evaluated as in~\cite{CLOUGH1989229}
\begin{align}
    G_{k,l}\left(f, T, \phi, p\right) = \frac{f}{f_{k,l}} \, \frac{\tanh\left( \frac{h\,c\,f}{2\,k_b\,T}\right)}{\tanh\left( \frac{h\,c\,f_{k,l}}{2\,k_b\,T}\right)}\,F_{k,l}\left(f\right),
\end{align}
with $h$ and $k_b$ respectively being the Plank and Bolzmann constants, $f_{k,l}$ representing the resonant frequency for the isotopologue $k$ of the $l-$th gas, and $F_{k,l}\left(f\right)$ denoting the Van Vleck-Weisskopf asymmetric line shape that can be calculated as in~\cite{RevModPhys.17.227}
\begin{align}
   & F_{k,l}\left(f\right) = 100\,c\frac{\alpha_{k,l}}{\pi}\frac{f}{f_{k,l}} \nonumber \\ & \times \left( \frac{1}{\left(f-f_{k,l}\right)^2+\alpha_{k,l}^2} + \frac{1}{\left(f+f_{k,l}\right)^2+\alpha_{k,l}^2} \right),
\end{align}
where $\alpha_{k,l}$ is the Lorentz half-width that can be expressed~as
\begin{align}
    \alpha_{k,l} = \left( \left(1-q_{k,l}\right)\alpha_o^{a} + q_{k,l}\,a_{k,l}^{o}\right) \frac{p}{p_o} \left(\frac{T_o}{T}\right)^{t},
\end{align}
with $t$ being the temperature broadening coefficient, while $\alpha_o^{a}$ and $a_{k,l}^{o}$ respectively denoting Lorentz half-width of air and the reference value of the Lorentz half-width for the isotopologue $k$ of the $l-$th gas. Note that $t$, $\alpha_o^{a}$, and $a_{k,l}^{o}$ can be directly extracted from HITRAN. 

Note that except from the aforementioned model for the molecular absorption coefficient, a number of approximations including~\cite{8568124,8580934,8123513} have been published. Although these approximations are very accurate and have been widely used, since their application is in the area of $100-500\,\rm{GHz}$, in this paper, we decided to use a more general model that can cover all the THz band. 

Finally, $h_{f}$ denotes the fog gain, which based on~\cite{itu}, can be analyzed~as
\begin{align}
    h_f = 10^{-\frac{1}{2}\kappa_f\,M\,\frac{d_1+d_2}{1000}},
    \label{Eq:h_f}
\end{align}
where $M$ is the liquid water density in the fog and $\kappa_f$ is the fog attenuation coefficient, which can be obtained~as
\begin{align}
    \kappa_f = \frac{0.819\,f}{\epsilon_i\left(1+\eta^2\right)},
\end{align}
with 
\begin{align}
    \eta = \frac{2+\epsilon_i}{\epsilon_r}.
    \label{Eq:eta}
\end{align}
In~\eqref{Eq:eta}, $\epsilon_r$ and $\epsilon_i$ are the real and imaginary part of the dielectric permittivity of water, which can be expressed~as
\begin{align}
    \epsilon_r = \frac{\epsilon_0-\epsilon_1}{1+\left(\frac{f}{f_p}\right)^2} + \frac{\epsilon_1-\epsilon_2}{1+\left(\frac{f}{f_s}\right)^2}+\epsilon_2
\end{align}
and
\begin{align}
    \epsilon_i = \frac{f}{f_p} \frac{\epsilon_0-\epsilon_1}{1+\left(\frac{f}{f_p}\right)^2} + \frac{f}{f_s} \frac{\epsilon_1-\epsilon_2}{1+\left(\frac{f}{f_s}\right)^2},
\end{align}
respectively. The parameters $f_p$ and $f_s$ denote the principal and secondary relaxation frequencies, which can be evaluated~as
\begin{align}
    f_p =& 20.2\times 10^{-9} - 146\times 10^{-9}\left(\theta-1\right) 
    \nonumber \\ & 
    + 316\times 10^{-9}\left(\theta-1\right)^2
\end{align}
and 
\begin{align}
    f_s = 39.9\,f_p,
\end{align}
while 
\begin{align}
    \epsilon_0 &= 77.66 + 103.3\,\left(\theta-1\right),\\
    \epsilon_1 &= 0.0671\,\epsilon_0, \\
    \epsilon_2 &= 3.52
\end{align}
and
\begin{align}
    \theta = \frac{300}{T}. 
\end{align}

As discussed in~\cite{9714471}, $h_i$ with $i=1,2$, follows a logarithmic distribution with probability density function (PDF) and cumulative distribution function (CDF) that can be, respectively, obtained~as
\begin{align}
    f_{h_i}(x) = \frac{\zeta_i^{k_i}}{\Gamma(k_i)} x^{k_i-1} \left(\ln\left(\frac{1}{x}\right)\right)^{k_i-1},\, \text{with } 0<x\leq 1
    \label{Eq:f_h_i}
\end{align}
and
\begin{align}
    F_{h_i}(x) = \frac{\Gamma\left(k_i,\zeta_i\,\ln\left(\frac{1}{x}\right)\right)}{\Gamma(k_i)},\quad\quad\quad\quad \text{with } 0<x\leq 1,
    \label{Eq:F_h_i}
\end{align}
where 
\begin{align}
    \zeta_i = \frac{4.343}{\beta_i\,d_i}.
    \label{Eq:zeta}
\end{align}
In~\eqref{Eq:f_h_i}--\eqref{Eq:zeta},  $k_1$ and $\beta_1$ stands for the foggy conditions of the TX-HRIS link, while $k_2$ and $\beta_2$ stands for the foggy conditions of the HRIS-RX link.  For instance, $(k_i, \beta_i)$, with $i\in\{1, 2\}$, refers to light fog, if $(2.32, 13.12)$, to moderate fog, if $(5.49, 12.06)$, to thick fog, if $(6, 23)$, and to dense fog, if $(36.06, 11.91)$.  

According to~\cite{B:Schenk-book}, the distortion noise due to the TX hardware imperfections can be modeled as a zero-mean complex Gaussian random process with variance that can be obtained~as
\begin{align}
    \sigma_{n_t}^2 = \kappa_t^2 \, P_s,
\end{align}
where $\kappa_t$ is the TX error vector magnitude (EVM) and $P_s$ is the transmission power. 
Similarly, for a given channel realization, the distortion noise due to the RX hardware imperfections can be modeled as a zero-mean complex Gaussian random process with variance that can be expressed as 
\begin{align}
    \sigma_{n_r}^2 = \kappa_r^2 \, h_g^2\,h_1^2\,h_2^2\, P_s, 
\end{align}
where $\kappa_r$ is the RX EVM. Of note, the EVM in THz wireless systems is in the range of $[0.07, 0.4]$~\cite{6630485,PhD:Boulogeorgos,Koenig2013,Boulogeorgos2020b,Boulogeorgos2019,8885655}.

\section{Performance Analysis}

This section is devoted to analyze the performance of HRIS-empowered THz wireless systems. In this direction, Section~\ref{SS:GL} provides a closed-form expression for the HRIS-empowered THz wireless system geometric losses. Section~\ref{SS:Statistical} presents the statistical characterization of the end-to-end channel in terms of PDF and CDF. The outage probability of the system is presented in Section~\ref{SS:OP}, while the throughput is given in Section~\ref{SS:Throughput}.  

\subsection{Geometric Losses}\label{SS:GL}

From~\eqref{Eq:hg}, it becomes evident that the geometric losses can be obtained~as
\begin{align}
    P_L = \frac{1}{h_g^2},
\end{align}
or equivalently
\begin{align}
    P_L = \frac{1}{h_{g,f}^2}\,\frac{1}{h_{g,m}^2}\,\frac{1}{h_f^2}.
\end{align}
Thus, from~\eqref{Eq:h_gf},~\eqref{Eq:h_gm}, and~\eqref{Eq:h_f}, we observe that the end-to-end geometric loss depends on the relative position of the HRIS in respect to the TX, the HRIS size, the TX and RX antenna gains, the transmission frequency, as well as the atmospheric conditions, i.e. temperature and liquid water density. 

\subsection{Statistical Characterization of the End-to-End Channel}\label{SS:Statistical}
Let 
\begin{align}
    A = h_1\,h_2,
\end{align}
the following theorem returns the PDF of $A$. 

\begin{thm}
    The PDF and CDF of $A$ can be, respectively, expressed~as
    \begin{align}
        f_{A}(x) &= \frac{\zeta_1^{k_1}\,\zeta_2^{k_2}}{\Gamma\left(k_1+k_2\right)}\,x^{\zeta_2-1}\,\left(-\ln\left(x\right)\right)^{k_1+k_2-1}
        \nonumber \\ & \times\,_1F_1\left(k_1;k_1+k_2;(\zeta_1-\zeta_2)\ln(x)\right) 
        \label{Eq:f_A}
    \end{align}
    and 
    \begin{align}
    &F_A(x)=(-1)^{k_1+k_2=1}\frac{\zeta_1^{k_1}\,\zeta_2^{k_2}}{k_1+k_2}\,
    \frac{\left(\ln\left(x\right)\right)^{k_1+k_2}}{\Gamma\left(k_1\right) \Gamma\left(k_2\right)}
    \nonumber \\ & \times
    \rm{H}_{A}\left(k_1+k_2, k_1; k_1+k_2+1; \left(\zeta_1-\zeta_2\right)\ln(x),\zeta_2\ln(x)\right),
    \label{Eq:F_A}
\end{align}
    with $x\in(0,1]$.
\end{thm}
\begin{IEEEproof}
    For brevity, the proof of Theorem 1 is given in Appendix A. 
\end{IEEEproof}

The PDF of the end-to-end channel coefficient
\begin{align}
    A_t = h_g\,A,
\end{align}
can be evaluated~as
\begin{align}
    f_{A_{t}}(x) = \frac{1}{h_{g}}\, f_{A}\left(\frac{x}{h_g}\right).
    \label{Eq:f_A_t}
\end{align}
By applying~\eqref{Eq:f_A} to~\eqref{Eq:f_A_t}, we obtain
\begin{align}
    f_{A_t}(x) &= \frac{\zeta_1^{k_1}\,\zeta_2^{k_2}}{h_g\Gamma\left(k_1+k_2\right)}\,\left(\frac{x}{h_g}\right)^{\zeta_2-1}\,\left(-\ln\left(\frac{x}{h_g}\right)\right)^{k_1+k_2-1}
        \nonumber \\ & \times\,_1F_1\left(k_1;k_1+k_2;(\zeta_1-\zeta_2)\ln(\frac{x}{h_g})\right). 
        \label{Eq:f_At}
\end{align}
Additionally, the CDF of $A_t$ can be expressed~as
\begin{align}
    F_{A_t}(x) = F_{A}\left(\frac{x}{h_g}\right),
\end{align}
which, by applying~\eqref{Eq:F_A}, yields
\begin{align}
    &F_{A_t}(x)=(-1)^{k_1+k_2=1}\frac{\zeta_1^{k_1}\,\zeta_2^{k_2}}{k_1+k_2}\,
    \frac{\left(\ln\left(\frac{x}{h_g}\right)\right)^{k_1+k_2}}{\Gamma\left(k_1\right) \Gamma\left(k_2\right)}
    \nonumber \\ & \times
    \rm{H}_{A}\left(k_1+k_2, k_1; k_1+k_2+1; \left(\zeta_1-\zeta_2\right)\ln\left(\frac{x}{h_g}\right),
   \right. \nonumber \\ & \left. 
    \zeta_2\ln\left(\frac{x}{h_g})\right)\right).
\end{align}

\subsection{Outage Probability} \label{SS:OP}

From~\eqref{Eq:r}, the signal-to-distortion-plus-noise-ratio (SDNR) can be obtained~as
\begin{align}
    \gamma = \frac{h_g^2\,A^2\,P_s}{h_g^2\,A^{2}\left(\kappa_t^2+\kappa_r^2\right)\,P_s + \sigma_n^2},
\end{align}
or equivalently
\begin{align}
    \gamma = \frac{A^2}{A^2\left(\kappa_t^2+\kappa_r^2\right)+\frac{1}{\rho}},
    \label{Eq:gamma}
\end{align}
where
\begin{align}
    \rho = \frac{h_g^2\,P_s}{\sigma_n^2}.
\end{align}

The outage probability is defined~as
\begin{align}
    P_o\left(\gamma_{\rm{th}}\right) = \Pr\left(\gamma\leq\gamma_{\rm{th}}\right),
    \label{Eq:Po_def}
\end{align}
where $\gamma_{\rm{th}}$ stands for the SNR threshold.
The following proposition returns a closed-form expression for the outage probability.
\begin{prop}
    The outage probability can be evaluated~as in~\eqref{Eq:P_o_final}, given at the top of the next page. 
\begin{figure*}
\begin{align}
    &P_o\left(\gamma_{\rm{th}}\right) = (-1)^{k_1+k_2=1}\frac{\zeta_1^{k_1}\,\zeta_2^{k_2}}{k_1+k_2}\,
    \frac{\left(\ln\left(\sqrt{\frac{\gamma_{\rm{th}}}{\rho}}\,\frac{1}{\sqrt{1-\gamma_{\rm{th}}\left(\kappa_t^2+\kappa_r^2\right)}}\right)\right)^{k_1+k_2}}{\Gamma\left(k_1\right) \Gamma\left(k_2\right)}
    \nonumber \\ & \times
    \rm{H}_{A}\left(k_1+k_2, k_1; k_1+k_2+1; \left(\zeta_1-\zeta_2\right)\ln\left(\sqrt{\frac{\gamma_{\rm{th}}}{\rho}}\,\frac{1}{\sqrt{1-\gamma_{\rm{th}}\left(\kappa_t^2+\kappa_r^2\right)}}\right),\zeta_2\ln\left(\sqrt{\frac{\gamma_{\rm{th}}}{\rho}}\,\frac{1}{\sqrt{1-\gamma_{\rm{th}}\left(\kappa_t^2+\kappa_r^2\right)}}\right)\right)
    \label{Eq:P_o_final}
\end{align}
\hrulefill
\end{figure*}
In~\eqref{Eq:P_o_final}, the following condition should hold:
\begin{align}
    \gamma_{\rm{th}}\leq \min\left(\frac{1}{\kappa_t^{2}+\kappa_r^2}, \frac{\rho}{\left(\kappa_t^2+\kappa_r^2\right)\rho+1}\right).
    \label{Eq:gamma_th_inequality}
\end{align}
Otherwise, $P_o\left(\gamma_{\rm{th}}\right)=1$.
\end{prop}
\begin{IEEEproof}
    For brevity, the proof of Theorem 1 is given in Appendix B. 
\end{IEEEproof}

\subsection{Throughput}\label{SS:Throughput}

The throughput is defined as
\begin{align}
    D = W\, r_{t}\,\left(1-P_o\left(\gamma_{\rm{th}}\right)\right),
    \label{Eq:D}
\end{align}
where $W$ and $r_{t}$ are respectively the bandwidth and the spectral efficiency of the modulation and coding scheme that is used. Note that $r_t$ and $\gamma_{\rm{th}}$ are connected through
\begin{align}
    r_t = \log_2\left(1+\gamma_{\rm{th}}\right).
\end{align}
As a result,~\eqref{Eq:D} can be rewritten~as
\begin{align}
    D = W\, r_{t}\,\left(1-P_o\left(2^{r_t}-1\right)\right).
    \label{Eq:D1}
\end{align}
By applying~\eqref{Eq:P_o_final} to~\eqref{Eq:D1}, we obtain~\eqref{Eq:D2}, given at the top of the next page. 
\begin{figure*}
\begin{align}
    D = W\, r_{t}\,&\left(1- 
    (-1)^{k_1+k_2=1}\frac{\zeta_1^{k_1}\,\zeta_2^{k_2}}{k_1+k_2}\,
    \frac{\left(\ln\left(\sqrt{\frac{2^{r_t}-1}{\rho}}\,\frac{1}{\sqrt{1-\left(2^{r_t}-1\right)\left(\kappa_t^2+\kappa_r^2\right)}}\right)\right)^{k_1+k_2}}{\Gamma\left(k_1\right) \Gamma\left(k_2\right)} \right.
    \nonumber \\ & \left. \times
    \rm{H}_{A}\left(k_1+k_2, k_1; k_1+k_2+1; 
    \left(\zeta_1-\zeta_2\right)\ln(\sqrt{\frac{\left(2^{r_t}-1\right)}{\rho}}\,\frac{1}{\sqrt{1-\left(2^{r_t}-1\right)\left(\kappa_t^2+\kappa_r^2\right)}}),
    \right.\right. \nonumber \\ & \left.\left.
    \zeta_2\ln(\sqrt{\frac{\left(2^{r_t}-1\right)}{\rho}}\,\frac{1}{\sqrt{1-\left(2^{r_t}-1\right)\left(\kappa_t^2+\kappa_r^2\right)}})\right) \right)
    \label{Eq:D2}
\end{align}
\hrulefill
\end{figure*}

Notice that~\eqref{Eq:D2} is valid, if and only if~\eqref{Eq:gamma_th_inequality} is satisfied. Or equivalently, if
\begin{align}
    r_{\rm{th}}\leq & \min\left(\log_2\left(\frac{1}{\kappa_t^{2}+\kappa_r^2}+1\right), \right.
    \nonumber \\ & \left.\log_2\left(1+\frac{\rho}{\left(\kappa_t^2+\kappa_r^2\right)\rho+1}\right)\right).
    \label{Eq:r_th_inequality}
\end{align}
Otherwise, the throughput is equal to $0$. 

\section{Results \& Discussion}

This section focuses on  verifying the  theoretical framework by means of  Monte Carlo simulations and extracting engineering insights. In what follows, lines stand for analytical results, while markers are used for simulations. Unless otherwise stated, black, red, orange, and blue colors are used for light, moderate, thick, and dense fog, respectively. 

\begin{figure}
	\centering
	\scalebox{1.00}{\input{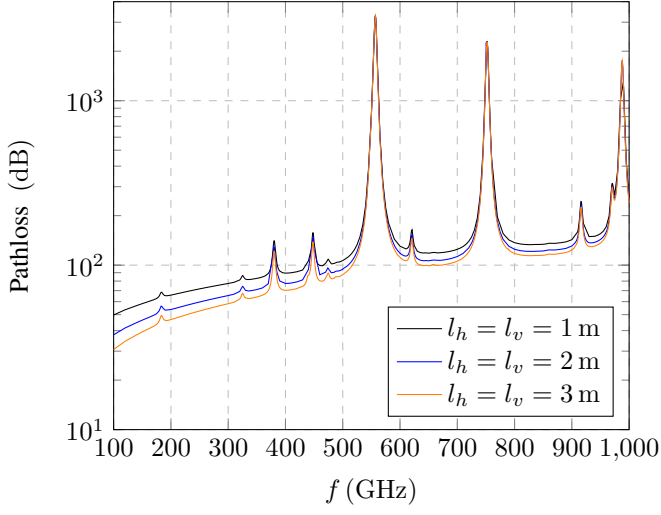}}
	\vspace{-0.25cm}
	\caption{Geometric loss as a function of $f$ for different values of $l_h=l_v$.}
	\label{Fig:PL_vs_f}
\end{figure}

Fig.~\ref{Fig:PL_vs_f} depicts the geometric loss as a function of $f$ for different values of $l_h=l_v$, assuming $G_{t}=G_r=50\,\rm{dBi}$, $\psi=\pi/4$, $T=20^{o}\rm{C}$, $P=101300.0\,\rm{Pa}$, $d_1=d_2=50\,\rm{m}$ and water vapour density equals to $7.5\,\rm{g/m^3}$. Notice that a water vapour density that is equal to $7.5\,\rm{g/m^2}$ corresponds to moderate fog.  From this figure, we observe that in the range of $100$ to $1000\,\rm{GHz}$, $10$ transmission windows exist. In more detail, the first transmission windows is from $100$ to $180\,\rm{GHz}$; there, the available bandwidth is equal to $80\,\rm{GHz}$. The second transmission window is from $192$ to $322\,\rm{GHz}$. The available bandwidth of the second transmission window is $130\,\rm{GHz}$. Notice that although the available bandwidth of the second transmission window is greater than the one of the first transmission window, the average geometric loss in the second transmission window is also higher than the one of the first  transmission window. The range of the third transmission window is $335-373\,\rm{GHz}$. As a consequence, the available bandwidth of the third transmission window is $38\,\rm{GHz}$. The forth transmission window is from $390$ to approximately $443\,\rm{GHz}$. Thus, the available bandwidth of the forth transmission window is $53\,\rm{GHz}$. The fifth transmission window is from $456$ to $472\,\rm{GHz}$, while the sixth transmission window is from $477$ to $487\,\rm{GHz}$. The seventh transmission window is from $587$ to $616\,\rm{GHz}$. The range of the eighth transmission window is from $630$ to $715\,\rm{GHz}$. The ninth transmission window is from $790$ to $890\,\rm{GHz}$, while the tenth is from $950$ to $970\,\rm{GHz}$. Finally, from a given $f$, as $l_h=l_v$ increases, i.e., as the size of the HRIS increases, the geometric loss decreases. 

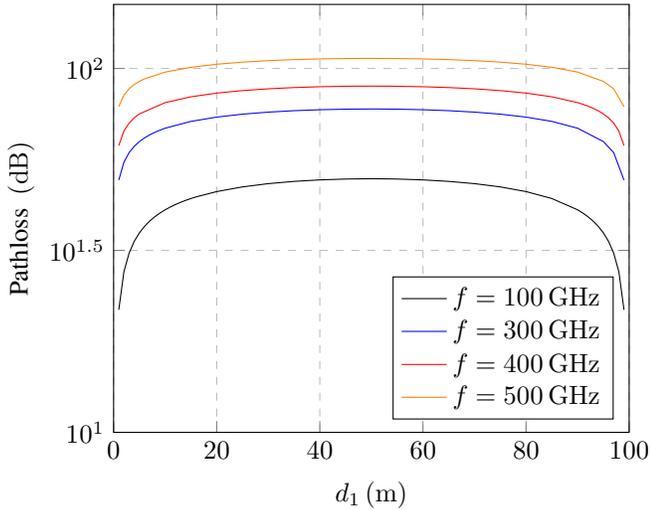
\begin{figure}
	\centering
	\scalebox{1.00}{\begin{tikzpicture}
\begin{axis}[
	xlabel={$d_1\,(\rm{m})$},
	ylabel={Pathloss $\,(\rm{dB})$},
    ymode = log,
	legend pos=south east,
	xmin = 0,
	xmax = 100,
	ymin = 10.0,
	ymax= 150,
	ymajorgrids=true,
	xmajorgrids=true,
	grid style=dashed,
	]

  \addplot[
	color=black]
	coordinates {
        (1, 21.7183)
        (2, 27.6508)
        (3, 31.0835)
        (4, 33.4923)
        (5, 35.3395)
        (6, 36.8312)
        (7, 38.0773)
        (8, 39.1432)
        (9, 40.0713)
        (10, 40.8905)
        (11, 41.6213)
        (12, 42.2789)
        (13, 42.8749)
        (14, 43.4182)
        (15, 43.9158)
        (16, 44.3736)
        (20, 45.8880)
        (25, 47.2657)
        (30, 48.2500)
        (35, 48.9453)
        (40, 49.4099)
        (45, 49.6771)
        (49, 49.7610)
        (50, 49.7644)
        (51, 49.7610)
        (52, 49.7505)
        (53, 49.7331)
        (55, 49.6771)
        (60, 49.4099)
        (65, 48.9453)
        (70, 48.2500)
        (75, 47.2657)
        (80, 45.8880)
        (85, 43.9158)
        (90, 40.8905)
        (91, 40.0713)
        (92, 39.1432)
        (93, 38.0773)
        (94, 36.8312)
        (95, 35.3395)
        (96, 33.4923)
        (97, 31.0835)
        (98, 27.6508)
        (99, 21.7183)
    };
    \addlegendentry{$f=100\,\rm{GHz}$}

    \addplot[
	color=blue]
	coordinates {
        (1, 49.3515)
        (2, 55.2840)
        (3, 58.7167)
        (4, 61.1255)
        (5, 62.9727)
        (6, 64.4644)
        (7, 65.7105)
        (8, 66.7764)
        (9, 67.7045)
        (10, 68.5237)
        (15, 71.5490)
        (20, 73.5212)
        (25, 74.8989)
        (30, 75.8832)
        (35, 76.5785)
        (40, 77.0431)
        (45, 77.3103)
        (50, 77.3976)
        (51, 77.3942)
        (55, 77.3103)
        (60, 77.0431)
        (65, 76.5785)
        (70, 75.8832)
        (75, 74.8989)
        (80, 73.5212)
        (85, 71.5490)
        (90, 68.5237)
        (95, 62.9727)
        (97, 58.7167)
        (99, 49.3515)
    };
    \addlegendentry{$f=300\,\rm{GHz}$}

    \addplot[
	color=red]
	coordinates {
        (1, 61.4010)
        (2, 67.3335)
        (3, 70.7662)
        (4, 73.1750)
        (5, 75.0222)
        (10, 80.5732)
        (15, 83.5985)
        (20, 85.5707)
        (25, 86.9484)
        (30, 87.9327)
        (35, 88.6280)
        (40, 89.0926)
        (45, 89.3598)
        (46, 89.3914)
        (47, 89.4158)
        (48, 89.4332)
        (49, 89.4437)
        (50, 89.4471)
        (51, 89.4437)
        (52, 89.4332)
        (53, 89.4158)
        (54, 89.3914)
        (55, 89.3598)
        (60, 89.0926)
        (65, 88.6280)
        (70, 87.9327)
        (75, 86.9484)
        (80, 85.5707)
        (85, 83.5985)
        (90, 80.5732)
        (91, 79.7540)
        (92, 78.8259)
        (93, 77.7600)
        (94, 76.5139)
        (95, 75.0222)
        (96, 73.1750)
        (97, 70.7662)
        (98, 67.3335)
        (99, 61.4010)
    };
    \addlegendentry{$f=400\,\rm{GHz}$}

    \addplot[
	color=orange]
	coordinates {
        (1, 78.5606)
        (2, 84.4930)
        (3, 87.9257)
        (4, 90.3345)
        (5, 92.1817)
        (6, 93.6735)
        (10, 97.7327)
        (15, 100.7581)
        (20, 102.7303)
        (25, 104.1079)
        (30, 105.0923)
        (35, 105.7875)
        (40, 106.2521)
        (45, 106.5194)
        (47, 106.5753)
        (48, 106.5928)
        (49, 106.6032)
        (50, 106.6067)
        (55, 106.5194)
        (60, 106.2521)
        (65, 105.7875)
        (70, 105.0923)
        (75, 104.1079)
        (80, 102.7303)
        (85, 100.7581)
        (90, 97.7327)
        (95, 92.1817)
        (97, 87.9257)
        (99, 78.5606)
    };
    \addlegendentry{$f=500\,\rm{GHz}$}

    \end{axis}
\end{tikzpicture}}
	\vspace{-0.25cm}
	\caption{Geometric loss vs $d_1$ for different values of $f$.}
	\label{Fig:PL_vs_d1}
\end{figure}
Fig.~\ref{Fig:PL_vs_d1} demonstrates the geometric loss as a function of $d_1$  for different values of $f$, assuming $d_1+d_2=100\,\rm{m}$, $G_{t}=G_r=50\,\rm{dBi}$, $l_h=l_v=1\,\rm{m}$, $\psi=\pi/4$, $T=20^{o}\rm{C}$, $P=101300.0\,\rm{Pa}$, and water vapour density equals to $7.5\,\rm{g/m^3}$. Note that the selected frequencies are within the transmission windows. From this figure, we observe that for fixed frequency and $d_1<\frac{d_1+d_2}{2}$, as $d_1$ increases, the geometric loss also increases, while for $d_1>\frac{d_1+d_2}{2}$, as $d_1$ increases, the geometric loss decreases. Likewise, for a given $f$, the maximum geometric loss is at $d_1=\frac{d_1+d_2}{2}$. For example, for $f=100\,\rm{GHz}$, as $d_1$ increases from $1$ to $10\,\rm{m}$, the geometric loss increases from $21.72$ to $40.89\,\rm{dB}$, while, for the same $f$, as $d_1$ increases from $90$ to $99\,\rm{m}$, the geometric loss increases from $40.89$ to $21.72\,\rm{dB}$. From this example, it becomes evident that for a given frequency, the same geometric loss is achieved for $d_1$ and $d_t-d_1$, where $d_t$ is the total distance. Moreover, we observe that based on the geometric loss, the optimal position of the HRIS is as near to the TX or the RX as possible.  Finally, for a given $d_1$, as $f$ increases, the geometric loss increases. 

\begin{figure}
	\centering
	\scalebox{1.00}{\input{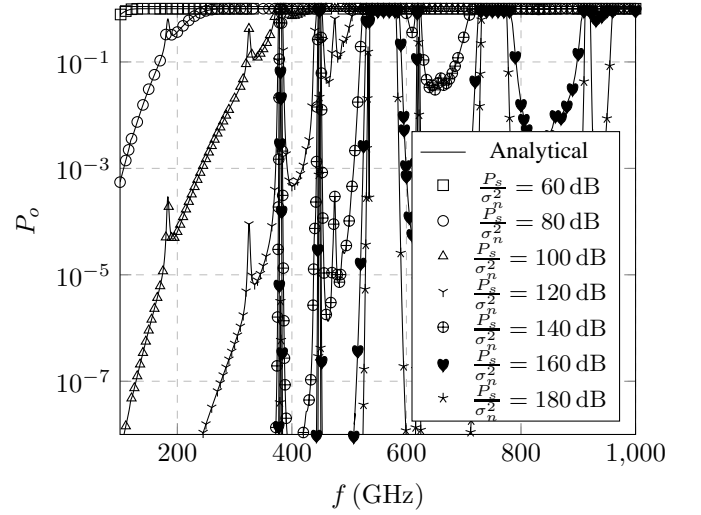}}
	\vspace{-0.25cm}
	\caption{Outage probability vs of $f$ for different values of $\frac{P_s}{\sigma_n^2}$.}
	\label{Fig:OP_vs_f}
\end{figure}

In Fig.~\ref{Fig:OP_vs_f}, the outage probability is plotted as a function of $f$, for different values of $\frac{P_s}{\sigma_n^2}$, assuming  $G_{t}=G_r=50\,\rm{dBi}$, $\psi=\pi/4$, $T=20^{o}\rm{C}$, $P=101300.0\,\rm{Pa}$, $d_1=d_2=50\,\rm{m}$,  $l_h=l_v=1\,\rm{m}$, and ideal RF front-end at both the TX and the RX. Moreover, the water vapour density is set to $7.5\,\rm{g/m^3}$. Notice that these conditions corresponds to moderate fog; thus, $k_1=k_2=5.49$ and $\beta_1=\beta+2=12.06$. As expected, for a given $f$, as $\frac{P_{s}}{\sigma_n^2}$ increases, the outage performance improves. For example, for $f=200\,\rm{GHz}$, the outage probability decreases from $1$ to $0.36$, as $\frac{P_{s}}{\sigma_n^2}$ increases from $60$ to $80\,\rm{dB}$. 
Additionally, for a fixed $\frac{P_{s}}{\sigma_n^2}$, a local maximum in the outage probability is observed, at the water resonating frequencies. 
Likewise, from this figure, we observe that based on the application reliability requirement, i.e., the maximum allowed outage probability, as well as the required energy consumption, which is translated to $\frac{P_{s}}{\sigma_n^2}$, a different range of frequencies can be employed. For instance, if the maximum allowed outage probability is set to $10^{-6}$, which is a realistic value for backhauling scenarios, and the transmission SNR is set to $100\,\rm{dB}$, the transmission frequency should be in the range of $[100, 151.5\,\rm{GHz}]$. Note that this frequency range is a subset of the first transmission frequency window, which is defined in Fig.~\ref{Fig:PL_vs_f}. For the same reliability requirement,  and for a transmission SNR equal to $120\,\rm{dB}$, we observe that the range of the transmission frequency increases to $[100, 309.5\,\rm{GHz}]$. In other words, as the transmission SNR increases, the range  of the transmission frequencies that can be used to support the system increases. As discussed in~\cite{Boulogeorgos2018}, this highlights the importance of adopting new type of transceivers in THz wireless systems that minimizes their noise figure. 

\begin{figure}
	\centering
	\scalebox{1.00}{\input{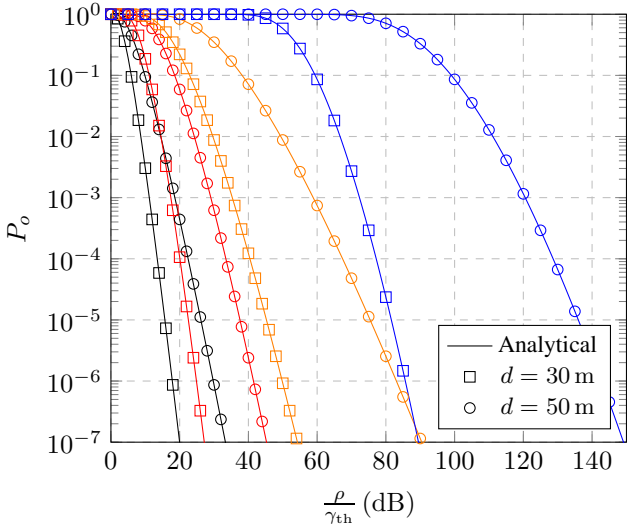}}
	\vspace{-0.25cm}
	\caption{Outage probability vs $\frac{\rho}{\gamma_{\rm{th}}}$ for different values of $d_1=d_2=d$, assuming light (black), moderate (red), thick (orange), and dense (blue) fog.}
	\label{Fig:OP_vs_r}
\end{figure}

Fig.~\ref{Fig:OP_vs_r} illustrates the outage probability as a function of $\frac{\rho}{\gamma_{\rm{th}}}$ for different values of $d_1=d_2=d$ and fog conditions. Moreover, we assume that the transceivers are equipped with ideal RF front-end, i.e. $\kappa_t=\kappa_r=0$. As expected for given fog conditions and $d$, as $\frac{\rho}{\gamma_{\rm{th}}}$ increases, the outage performance improves. For example for light fog and $d=30\,\rm{m}$, as $\frac{\rho}{\gamma_{\rm{th}}}$ increases from $5$ to $10\,\rm{dB}$, the outage probability decreases for about $2$ orders of magnitude. Moreover, for fixed fog conditions and $\frac{\rho}{\gamma_{\rm{th}}}$, as $d$ increases, the outage probability also increases. For instance, for light fog and  $\frac{\rho}{\gamma_{\rm{th}}}=15\,\rm{dB}$, the outage probability increases from $2.08\times 10^{-5}$ to $7.63\times 10^{-3}$, as $d$ increases from $30$ to $50\,\rm{m}$. Finally, for given $\frac{\rho}{\gamma_{\rm{th}}}$ and $d$, as the fog becomes more severe, the outage performance degrades. For example, for  $\frac{\rho}{\gamma_{\rm{th}}}=15\,\rm{dB}$ and $d=30\,\rm{m}$, as the fog conditions changes from light to moderate, the outage probability increases from $2.08\times 10^{-5}$ to $7.15\times 10^{-3}$. For the same  $\frac{\rho}{\gamma_{\rm{th}}}$ and $d$, the system under thick fog, achieves an outage probability that is equal to $5.9\times 10^{-1}$, while, under dense fog, it is equal to $1$. These examples indicate the importance of accounting for the fog impact on the performance of HRIS-empowered THz wireless systems.

\begin{figure}
	\centering
	\scalebox{1.00}{\input{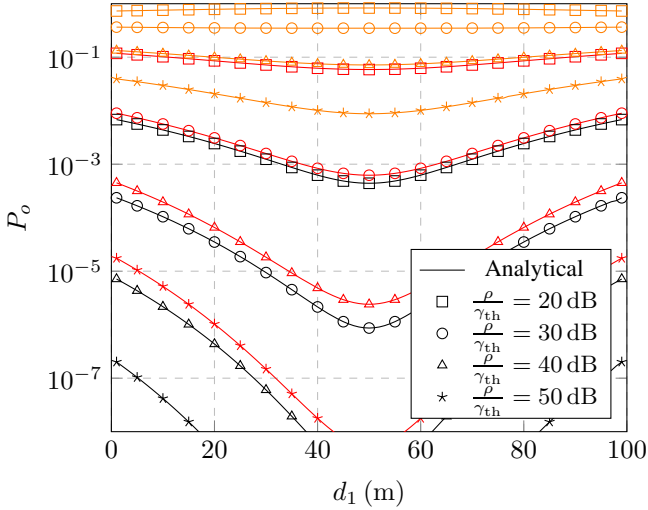}}
	\vspace{-0.25cm}
	\caption{Outage probability vs $d_1$ for different values of $\frac{\rho}{\gamma_{\rm{th}}}$, assuming light (black), moderate (red), thick (orange) fog and $d_1+d_2=100\,\rm{m}$.}
	\label{Fig:OP_vs_d1}
\end{figure}

Fig.~\ref{Fig:OP_vs_d1} depicts the outage probability as a function of $d_1$ for different values of $\frac{\rho}{\gamma_{\rm{th}}}$ and fog conditions, assuming $d_1+d_2=100\,\rm{m}$ and ideal RF front-end. From this figure, we observe that for given $\frac{\rho}{\gamma_{\rm{th}}}$ and fog conditions, for $d_1<\frac{d_1+d_2}{2}$, as $d$ increases the outage probability decreases. For example, for $\frac{\rho}{\gamma_{\rm{th}}}=40\,\rm{dB}$ and light fog, as $d_1$ increases from $10$ to $20\,\rm{m}$, the outage probability decreases from $2.17\times 10^{-6}$ to $4.35\times 10^{-7}$. On the other hand, for fixed $\frac{\rho}{\gamma_{\rm{th}}}$ and fog conditions, for $d_1>\frac{d_1+d_2}{2}$, as $d_1$ increases, the outage probability also increases. For instance, $\frac{\rho}{\gamma_{\rm{th}}}=40\,\rm{dB}$ and light fog, as $d_1$ increases from $80$ to $90\,\rm{m}$, the outage probability increases from  $4.35\times 10^{-7}$ to $2.17\times 10^{-6}$. For given fog conditions and $\frac{\rho}{\gamma_{\rm{th}}}$,the outage probability achieves the minimum value for $d_1=\frac{d_1+d_2}{2}$. In other words, the optimal place for the HRIS is at $\frac{d_1+d_2}{2}$. Moreover, we observe that the same outage performance are achieved for $d_1$ and $d_t-d_1$, where $d_t=d_1+d_2$. For example, for $\frac{\rho}{\gamma_{\rm{th}}}=40\,\rm{dB}$ and light fog, for both $d_1=10\,\rm{m}$ and $d_1=90\,\rm{m}$, the outage probability is equal to $2.17\times 10^{-6}$. Additionally, for given $d$ and fog conditions, as $\frac{\rho}{\gamma_{\rm{th}}}$ increases, the outage probability decreases. For instance, for $d=10\,\rm{m}$ and light fog, as $\frac{\rho}{\gamma_{\rm{th}}}$ changes from $20$ to $40\,\rm{dB}$, the outage probability decreases for approximately $5$ orders of magnitude. Finally, for fixed $d$ and $\frac{\rho}{\gamma_{\rm{th}}}$, as the fog density increases, the outage performance of the system degrades. 

\begin{figure}
	\centering
	\scalebox{1.00}{\input{images/OP_vs_kappa.tex}}
    (a) \\
    \scalebox{1.00}{\begin{tikzpicture}
\begin{axis}[
	xlabel={$\kappa$},
	ylabel={$P_o$},
    ymode = log,
	legend pos=north west,
	xmin = 0,
	xmax = 0.6,
	ymin = 10^-7,
	ymax= 10^0,
	ymajorgrids=true,
	xmajorgrids=true,
	grid style=dashed,
	]

 \addplot[
	color=red]
	coordinates {
        (0, 2.40469*10^-6)
        (0.05, 2.42015*10^-6)
        (0.07, 2.43511*10^-6)
        (0.10, 2.46734*10^-6)
        (0.12, 2.49562*10^-6)
        (0.15, 2.54885*10^-6)
        (0.17, 2.59202*10^-6)
        (0.20, 2.6693*10^-6)
        (0.22, 2.72991*10^-6)
        (0.25, 2.83598*10^-6)
        (0.27, 2.91787*10^-6)
        (0.30, 3.05975*10^-6)
        (0.32, 3.16863*10^-6)
        (0.35, 3.35675*10^-6)
        (0.37, 3.50109*10^-6)
        (0.40, 3.75105*10^-6)
        (0.42, 3.94364*10^-6)
        (0.45, 4.27929*10^-6)
        (0.47, 4.53998*10^-6)
        (0.50, 4.99894*10^-6)
        (0.52, 5.3596*10^-6)
        (0.55, 6.00363*10^-6)
        (0.57, 6.51781*10^-6)
        (0.60, 7.45353*10^-6)
    };

    \addplot[
	color=red,
    only marks,
    mark = square]
	coordinates {
        (0, 2.40469*10^-6)
        (0.05, 2.42015*10^-6)
        (0.07, 2.43511*10^-6)
        (0.10, 2.46734*10^-6)
        (0.12, 2.49562*10^-6)
        (0.15, 2.54885*10^-6)
        (0.17, 2.59202*10^-6)
        (0.20, 2.6693*10^-6)
        (0.22, 2.72991*10^-6)
        (0.25, 2.83598*10^-6)
        (0.27, 2.91787*10^-6)
        (0.30, 3.05975*10^-6)
        (0.32, 3.16863*10^-6)
        (0.35, 3.35675*10^-6)
        (0.37, 3.50109*10^-6)
        (0.40, 3.75105*10^-6)
        (0.42, 3.94364*10^-6)
        (0.45, 4.27929*10^-6)
        (0.47, 4.53998*10^-6)
        (0.50, 4.99894*10^-6)
        (0.52, 5.3596*10^-6)
        (0.55, 6.00363*10^-6)
        (0.57, 6.51781*10^-6)
        (0.60, 7.45353*10^-6)
    };

    \addplot[
	color=red]
	coordinates {
        (0.0, 0.000042393)
        (0.02, 0.0000425231)
        (0.05, 0.0000432149)
        (0.07, 0.0000440248)
        (0.10, 0.0000458174)
        (0.12, 0.0000474466)
        (0.15, 0.0000506631)
        (0.17, 0.0000534231)
        (0.20, 0.0000587292)
        (0.22, 0.0000632445)
        (0.25, 0.0000719807)
        (0.27, 0.0000795383)
        (0.30, 0.0000945807)
        (0.32, 0.000108085)
        (0.35, 0.000136384)
        (0.37, 0.000163431)
        (0.40, 0.000225214)
        (0.42, 0.000290819)
        (0.45, 0.000465651)
        (0.47, 0.000692023)
        (0.50, 0.00153889)
        (0.52, 0.00335909)
        (0.53, 0.00572699)
        (0.535, 0.00793978)
        (0.54, 0.0116767)
        (0.55, 0.0343192)
        (0.56, 0.369385)
        (0.561, 0.609609)
        (0.562, 0.98873)
        (0.5625, 1.0)
        (0.57, 1.0)
        (0.58, 1.0)
        (0.59, 1.0)
        (0.60, 1.0)
    };

    \addplot[
	color=red,
    only marks,
    mark = o]
	coordinates {
        (0.0, 0.000042393)
        (0.02, 0.0000425231)
        (0.05, 0.0000432149)
        (0.07, 0.0000440248)
        (0.10, 0.0000458174)
        (0.12, 0.0000474466)
        (0.15, 0.0000506631)
        (0.17, 0.0000534231)
        (0.20, 0.0000587292)
        (0.22, 0.0000632445)
        (0.25, 0.0000719807)
        (0.27, 0.0000795383)
        (0.30, 0.0000945807)
        (0.32, 0.000108085)
        (0.35, 0.000136384)
        (0.37, 0.000163431)
        (0.40, 0.000225214)
        (0.42, 0.000290819)
        (0.45, 0.000465651)
        (0.47, 0.000692023)
        (0.50, 0.00153889)
        (0.52, 0.00335909)
        (0.53, 0.00572699)
        (0.535, 0.00793978)
        (0.54, 0.0116767)
        (0.55, 0.0343192)
        (0.56, 0.369385)
        (0.561, 0.609609)
        (0.562, 0.98873)
        (0.5625, 1.0)
        (0.57, 1.0)
        (0.58, 1.0)
        (0.59, 1.0)
        (0.60, 1.0)
    };

    \addplot[
	color=red]
	coordinates {
        (0.00, 0.000621889)
        (0.02, 0.00062749)
        (0.05, 0.000658098)
        (0.07, 0.000695722)
        (0.10, 0.000786426) 
        (0.12, 0.000878704)
        (0.15, 0.00109343)
        (0.17, 0.0013196)
        (0.20, 0.00190263)
        (0.22, 0.00262044)
        (0.25, 0.0050746)
        (0.27, 0.00969845)
        (0.28, 0.0150016)
        (0.29, 0.0264247)
        (0.30, 0.0588796)
        (0.305, 0.104136)
        (0.31, 0.233654)
        (0.315, 0.871455)
        (0.3155, 0.974529)
        (0.316, 1.0)
        (0.317, 1.0)
        (0.32, 1.0)
        (0.33, 1.0)
        (0.34, 1.0)
        (0.35, 1.0)
        (0.37, 1.0)
        (0.40, 1.0)
        (0.42, 1.0)
        (0.45, 1.0)
        (0.47, 1.0)
        (0.50, 1.0)
        (0.52, 1.0)
        (0.55, 1.0)
        (0.57, 1.0)
        (0.60, 1.0)
    };

     \addplot[
	color=red,
    only marks,
    mark = triangle]
	coordinates {
        (0.00, 0.000621889)
        (0.02, 0.00062749)
        (0.05, 0.000658098)
        (0.07, 0.000695722)
        (0.10, 0.000786426) 
        (0.12, 0.000878704)
        (0.15, 0.00109343)
        (0.17, 0.0013196)
        (0.20, 0.00190263)
        (0.22, 0.00262044)
        (0.25, 0.0050746)
        (0.27, 0.00969845)
        (0.28, 0.0150016)
        (0.29, 0.0264247)
        (0.30, 0.0588796)
        (0.305, 0.104136)
        (0.31, 0.233654)
        (0.315, 0.871455)
        (0.3155, 0.974529)
        (0.316, 1.0)
        (0.317, 1.0)
        (0.32, 1.0)
        (0.33, 1.0)
        (0.34, 1.0)
        (0.35, 1.0)
        (0.37, 1.0)
        (0.40, 1.0)
        (0.42, 1.0)
        (0.45, 1.0)
        (0.47, 1.0)
        (0.50, 1.0)
        (0.52, 1.0)
        (0.55, 1.0)
        (0.57, 1.0)
        (0.60, 1.0)
    };
     
    \addplot[
	color=orange]
	coordinates {
        (0.00, 0.0716832)
        (0.02, 0.0717065)
        (0.05, 0.0718292)
        (0.07, 0.07197)
        (0.10, 0.072271)
        (0.12, 0.0725329)
        (0.15, 0.0730202)
        (0.17, 0.07341)
        (0.20, 0.0740965)
        (0.22, 0.0746251)
        (0.25, 0.0755302)
        (0.27, 0.0762126)
        (0.30, 0.0773631)
        (0.32, 0.0782204)
        (0.35, 0.0796532)
        (0.37, 0.0807139)
        (0.40, 0.0824788)
        (0.42, 0.0837819)
        (0.45, 0.0859474)
        (0.47, 0.0875463)
        (0.50, 0.0902071)
        (0.52, 0.0921767)
        (0.55, 0.095467)
        (0.57, 0.0979146)
        (0.60, 0.10203)
    };

    \addplot[
	color=orange,
    only marks,
    mark = square]
	coordinates {
        (0.00, 0.0716832)
        (0.02, 0.0717065)
        (0.05, 0.0718292)
        (0.07, 0.07197)
        (0.10, 0.072271)
        (0.12, 0.0725329)
        (0.15, 0.0730202)
        (0.17, 0.07341)
        (0.20, 0.0740965)
        (0.22, 0.0746251)
        (0.25, 0.0755302)
        (0.27, 0.0762126)
        (0.30, 0.0773631)
        (0.32, 0.0782204)
        (0.35, 0.0796532)
        (0.37, 0.0807139)
        (0.40, 0.0824788)
        (0.42, 0.0837819)
        (0.45, 0.0859474)
        (0.47, 0.0875463)
        (0.50, 0.0902071)
        (0.52, 0.0921767)
        (0.55, 0.095467)
        (0.57, 0.0979146)
        (0.60, 0.10203)
    };
     
     \addplot[
	color=orange]
	coordinates {
        (0.00, 0.170744)
        (0.02, 0.170894)
        (0.05, 0.171683)
        (0.07, 0.172594)
        (0.09, 0.173827)
        (0.10, 0.174567)
        (0.12, 0.17631)
        (0.15, 0.179621)
        (0.17, 0.182336)
        (0.20, 0.187269)
        (0.22, 0.191203)
        (0.25, 0.19824)
        (0.27, 0.203813)
        (0.30, 0.213786)
        (0.32, 0.221736)
        (0.35, 0.236155)
        (0.37, 0.247878)
        (0.40, 0.269767)
        (0.42, 0.288255)
        (0.45, 0.324742)
        (0.47, 0.357922)
        (0.50, 0.431587)
        (0.52, 0.511514)
        (0.55, 0.773586)
        (0.56, 0.975743)
        (0.561, 0.993465)
        (0.562, 0.999985)
        (0.565, 1.0)
        (0.57, 1.0)
        (0.58, 1.0)
        (0.59, 1.0)
        (0.60, 1.0)
    };

    \addplot[
	color=orange,
    only marks,
    mark = o]
	coordinates {
        (0.00, 0.170744)
        (0.02, 0.170894)
        (0.05, 0.171683)
        (0.07, 0.172594)
        (0.09, 0.173827)
        (0.10, 0.174567)
        (0.12, 0.17631)
        (0.15, 0.179621)
        (0.17, 0.182336)
        (0.20, 0.187269)
        (0.22, 0.191203)
        (0.25, 0.19824)
        (0.27, 0.203813)
        (0.30, 0.213786)
        (0.32, 0.221736)
        (0.35, 0.236155)
        (0.37, 0.247878)
        (0.40, 0.269767)
        (0.42, 0.288255)
        (0.45, 0.324742)
        (0.47, 0.357922)
        (0.50, 0.431587)
        (0.52, 0.511514)
        (0.55, 0.773586)
        (0.56, 0.975743)
        (0.561, 0.993465)
        (0.562, 0.999985)
        (0.565, 1.0)
        (0.57, 1.0)
        (0.58, 1.0)
        (0.59, 1.0)
        (0.60, 1.0)
    };

    \addplot[
	color=orange]
	coordinates {
    (0, 0.348751)
    (0.02, 0.349514)
    (0.05, 0.353588)
    (0.07, 0.358384)
    (0.10, 0.369111)
    (0.12, 0.379007)
    (0.15, 0.399011)
    (0.17, 0.416738)
    (0.20, 0.452584)
    (0.22, 0.485315)
    (0.23, 0.505437)
    (0.25, 0.556409)
    (0.27, 0.629475)
    (0.30, 0.83149)
    (0.31, 0.950278)
    (0.315, 0.999313)
    (0.32, 1.0)
    (0.35, 1.0)
    (0.37, 1.0)
    (0.40, 1.0)
    (0.42, 1.0)
    (0.45, 1.0)
    (0.47, 1.0)
    (0.48, 1.0)
    (0.50, 1.0)
    (0.52, 1.0)
    (0.55, 1.0)
    (0.57, 1.0)
    (0.60, 1.0)
    };

    \addplot[
	color=orange,
    only marks,
    mark = triangle]
	coordinates {
    (0, 0.348751)
    (0.02, 0.349514)
    (0.05, 0.353588)
    (0.07, 0.358384)
    (0.10, 0.369111)
    (0.12, 0.379007)
    (0.15, 0.399011)
    (0.17, 0.416738)
    (0.20, 0.452584)
    (0.22, 0.485315)
    (0.23, 0.505437)
    (0.25, 0.556409)
    (0.27, 0.629475)
    (0.30, 0.83149)
    (0.31, 0.950278)
    (0.315, 0.999313)
    (0.32, 1.0)
    (0.35, 1.0)
    (0.37, 1.0)
    (0.40, 1.0)
    (0.42, 1.0)
    (0.45, 1.0)
    (0.47, 1.0)
    (0.48, 1.0)
    (0.50, 1.0)
    (0.52, 1.0)
    (0.55, 1.0)
    (0.57, 1.0)
    (0.60, 1.0)
    };
    
     \end{axis}

\end{tikzpicture}} 
    (b) \\
	\caption{Outage probability vs $\kappa$ for different values of ${\gamma_{\rm{th}}}$, assuming light (black), moderate (red), thick (orange) fog and $\rho=30\,\rm{dB}$ (a) and $\rho=40\,\rm{dB}$ (b).}
	\label{Fig:OP_vs_kappa}
\end{figure}
Fig.~\ref{Fig:OP_vs_kappa} demonstrates the impact of transceiver hardware imperfections on the outage performance of HRIS-empowered THz wireless systems. In more detail, the outage probability is plotted as a function of $\kappa=\sqrt{\kappa_t^2+\kappa_r^2}$ for different values of $\gamma_{\rm{th}}$, $\rho$, as well as fog conditions, assuming $d_1=d_2=50\,\rm{m}$. As expected, for given  $\gamma_{\rm{th}}$, $\rho$ and fog conditions, as $\kappa$ increases, the impact of transceivers hardware imperfections increases; thus, the outage probability increases. For example, for $\gamma_{\rm{th}}=10\,\rm{dB}$, $\rho=30\,\rm{dB}$, and light fog, the outage probability increases for more than $2$ orders of magnitude, as $\kappa$ increases from $0.1$ to $0.3$. This indicates the importance of taking into account the impact of transceiver hardware imperfections, when designing modulation and coding schemes. From this figure, it become evident that, for given $\kappa$, $\rho$, and fog conditions, as $\gamma_{\rm{th}}$ increases, the impact of hardware imperfections becomes more severe; as a consequence, an outage performance degradation is observed. For instance, for $\kappa=0.2$, $\rho=30\,\rm{dB}$, and light fog, the outage probability increases for more than $1$ order of magnitude, as $\gamma_{\rm{th}}$ increases from $5$ to $10\,\rm{dB}$. Moreover, or given $\kappa$, $\rho$, and $\gamma_{\rm{th}}$, as the density of fog increases, the outage probability also increases. For example, for  $\kappa=0.2$, $\rho=30\,\rm{dB}$, and $\gamma_{\rm{th}}=5\,\rm{dB}$, as the fog conditions change from light to moderate, the outage probability increases for more than $2$ orders of magnitude.  Finally, for given  $\kappa$, $\gamma_{\rm{th}}$, and fog conditions, as $\rho$ increases, the outage probability decreases.

\begin{figure}
	\centering
	\scalebox{1.00}{\input{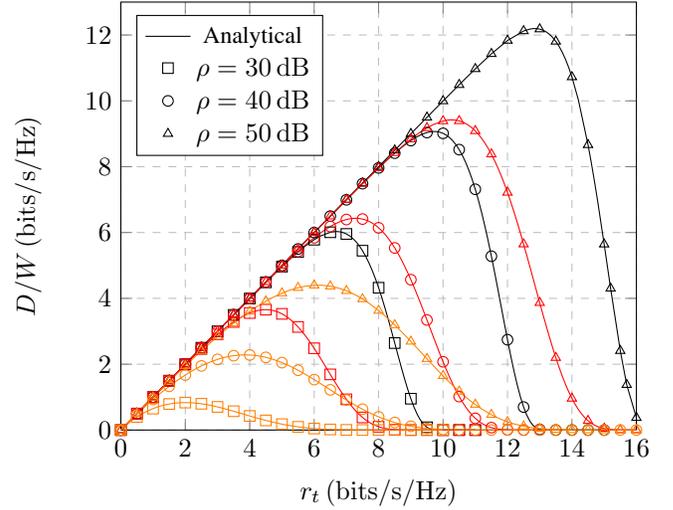}}
	\vspace{-0.25cm}
	\caption{$D/W$ vs $r_t$ for different values of ${\rho}$, assuming light (black), moderate (red), thick (orange) fog and $d_1=d_2=50\,\rm{m}$.}
	\label{Fig:Th_vs_rt}
\end{figure}
Fig.~\ref{Fig:Th_vs_rt} presents the throughput to bandwidth ratio as a function of $r_t$ for different values of ${\rho}$ and fog conditions, assuming ideal RF front-end and $d_1=d_2=50\,\rm{m}$. From this figure, we observe that for given $\rho$ and fog condition, an optimal spectral efficiency of the modulation and coding transmission scheme, $r_{t}^{o}$, exists that maximizes the throughput to bandwidth ratio. For $r_t<r_{t}^{o}$, as $r_{t}$ increases, $\frac{D}{W}$ also increases. On the other hand, for  $r_t>r_{t}^{o}$, as $r_{t}$ increases, $\frac{D}{W}$ decreases. Moreover, for fixed $\rho_t$ and fog conditions, as $\rho$ increases, $\frac{D}{W}$ also increases. For example, under light fog, for $r_t=8\,\rm{bits/s/Hz}$, the $D/W$ increases from $4.32$ to $7.96\,\rm{bits/s/Hz}$, as $\rho$ increases from $30$ to $40\,\rm{dB}$. Finally, for given $r_t$ and $\rho$, as the fog density increases, $D/W$ decreases.    

\begin{figure}
	\centering
	\scalebox{1.00}{\input{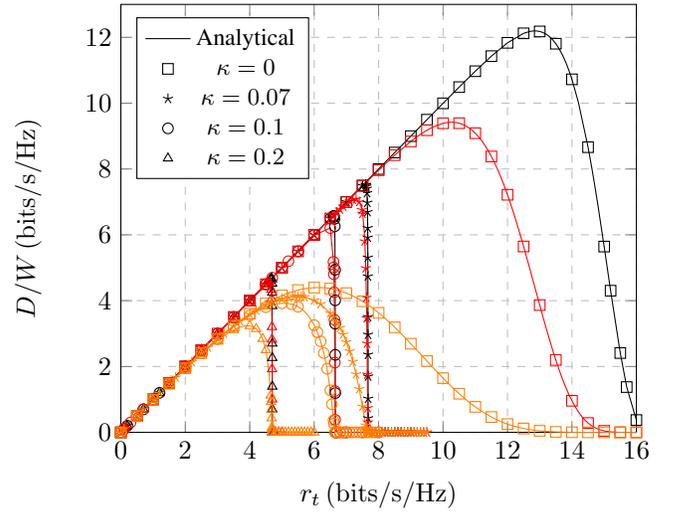}}
	\vspace{-0.25cm}
	\caption{$D/W$ vs $r_t$ for different values of ${\kappa}$, assuming light (black), moderate (red), thick (orange) fog, $d_1=d_2=50\,\rm{m}$ and $\rho=50\,\rm{dB}$.}
	\label{Fig:Th_vs_rt_k}
\end{figure}

Fig.~\ref{Fig:Th_vs_rt_k} presents the throughput to bandwidth ratio as a function of the transmission scheme spectral efficiency for different values of $\kappa$ and under different fog conditions. As a benchmark, the case of ideal RF front-end is plotted. As expected, an optimal transmission spectral efficiency, $r_{t}^{o}$, exists, for which the throughput to bandwidth ratio is maximized. For given $\kappa$, fog density, and $r_t<r_{t}^{o}$, as $r_{t}$ increases, $D/W$ also increases. For example, thick fog and $\kappa=0.07$, the optimal transmission spectral efficiency is equal to $5.5\,\rm{bits/s/Hz}$, where the achievable throughput to bandwidth ratio is $4.125\,\rm{bits/s/Hz}$. Moreover, as the transmission spectral efficiency increases from $4$ to $5\,\rm{bits/s/Hz}$, the throughput to bandwidth ratio increases from $3.58$ to $4.058\,\rm{bits/s/Hz}$. On the other hand, for $r_{t}>r_{t}^{o}$, as the transmission spectral efficiency increases,  the throughput to bandwidth ratio decreases. For instance, for thick fog and $\kappa=0.07$, as the transmission spectral efficiency increases from $6$ to $7\,\rm{bits/s/Hz}$, the throughput to bandwidth ratio decreases from $4.01$ to $2.74\,\rm{bits/s/Hz}$. Likewise, for given fog density and transmission spectral efficiency, as $\kappa$ increases,  the throughput to bandwidth ratio decreases. For example, for thick fog and transmission spectral efficiency equal to $6\,\rm{bits/s/Hz}$,  the throughput to bandwidth ratio decreases from  $4.01$ to $3.26\,\rm{bits/s/Hz}$, as $\kappa$ increases from $0.07$ to $0.1$. Additionally, for fixed transmission spectral efficiency and $\kappa$, as the fog density increases, the throughput to bandwidth ratio decreases. For instance, for $r_t=6\,\rm{bits/s/Hz}$ and $\kappa=0.07$, if the fog condition changes from light to thick, the throughput to bandwidth ratio will decrease from $6.0$ to $3.26\,\rm{bits/s/Hz}$. This indicates the importance of accurately modeling the fog conditions, when estimating the throughput performance of HRIS-empowered wireless THz systems. Finally, from this figure, it becomes apparent that a maximum transmission spectral efficiency exists beyond which the throughput to bandwidth ratio becomes equal to $0$. The maximum transmission spectral efficiency depends on the level of transceiver hardware imperfections. For example, for $\kappa=0.07$, it is equal to $7.68\,\rm{bits/s/Hz}$, while, for $\kappa=0.1$, it is equal to $6.66\,\rm{bits/s/Hz}$. Note that the maximum transmission spectral efficiency is independent of the fog conditions. Also, we observe that the maximum transmission spectral efficiency is lower than the optimal  transmission spectral efficiency for the case of ideal RF front-end. This highlights the importance of accounting for the impact of hardware imperfections, when selecting the transmission scheme. 


\begin{figure}
	\centering
	\scalebox{1.00}{\input{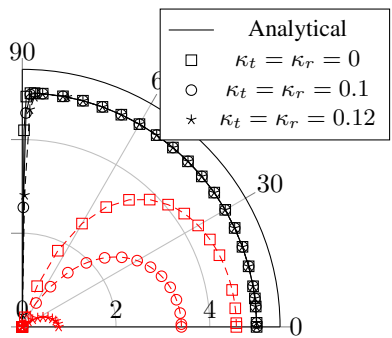}}
	\vspace{-0.25cm}
	\caption{$D/W$ vs $\psi$ for different values of $\kappa_t=\kappa_r$, assuming $f=100\,\rm{GHz}$ (black) and $f=370\,\rm{GHz}$ (red).}
	\label{Fig:D_vs_psi}
\end{figure}

In Fig.~\ref{Fig:D_vs_psi}, $D/W$ is illustrated as a function of $\psi$ for different levels of transceiver hardware imperfections and frequencies, assuming $\frac{P_s}{\sigma_n^2}=120\,\rm{dB}$, $r_t=5\,\rm{bits/s/Hz}$, $G_{t}=G_r=50\,\rm{dBi}$, $\psi=\pi/4$, $T=20^{o}\rm{C}$, $P=101300.0\,\rm{Pa}$, $d_1=d_2=50\,\rm{m}$,  $l_h=l_v=1\,\rm{m}$, and water vapor equal to $7.5\,\rm{g/m^3}$, which corresponds to moderate fog conditions. As a benchmark, the ideal RF front-end scenario is also plotted. From this figure, we observe that for given $\kappa_t=\kappa_r$ and $f$, as $\psi$ increases, $D/W$ decreases. For example, for $\kappa_t=\kappa_r=0$ and $f=370\,\rm{GHz}$, as $\psi$ increases from $10^o$ to $45^{o}$, $D/W$ decreases from $4.55$ to $3.84\,\rm{bits/s/Hz}$. This indicates that not only the relative distance from the TX and the RX, but also the orientation of the HRIS plays a pivotal role on the throughput performance of HRIS-empowered wireless THz systems. Additionally, it is become apparent that, for fixed $\psi$ and $\kappa_t=\kappa_r$, as the geometric loss increases through a frequency increase, $D/W$ decreases. For instance, for $\psi=45^{o}$ and $\kappa_t=\kappa_r=0$, $D/W$ decreases from $5.0$ to $3.84\,\rm{bits/s/Hz}$, as $f$ increases from $100$ to $370\,\rm{GHz}$. This example reveals the importance of appropriately allocating the frequencies in order to make the most out of the available bandwidth. Finally, for given $f$ and $\psi$, as the level of transceiver hardware imperfections increase, $D/W$ decreases.      

\begin{figure}
	\centering
	\scalebox{1.00}{\begin{tikzpicture}
\begin{axis}[
	xlabel={$d_1\,\rm{(m)}$},
	ylabel={$D/W\,\rm{(bits/s/Hz)}$},
	legend pos=south east,
	xmin = 0,
	xmax = 100,
	ymin = 0,
	ymax= 5.1,
	ymajorgrids=true,
	xmajorgrids=true,
	grid style=dashed,
	]

 \addplot[
	color=black]
	coordinates {
        (1, 4.99923)
        (5, 4.96028)
        (10, 4.83623)
        (20, 4.46218)
        (25, 4.25593)
        (30, 4.0572)
        (35, 3.88066)
        (40, 3.74113)
        (45, 3.6515)
        (50, 3.62052)
        (55, 3.6515)
        (60, 3.74113)
        (65, 3.88066)
        (70, 4.0572)
        (75, 4.25593)
        (80, 4.46218)
        (90, 4.83623)
        (95, 4.96028)
        (99, 4.99923)
 };
 \addlegendentry{\small{Analytical}}

\addplot[
	color=black,
    only marks,
    mark=square]
	coordinates {
        (1, 4.99923)
        (5, 4.96028)
        (10, 4.83623)
        (20, 4.46218)
        (25, 4.25593)
        (30, 4.0572)
        (35, 3.88066)
        (40, 3.74113)
        (45, 3.6515)
        (50, 3.62052)
        (55, 3.6515)
        (60, 3.74113)
        (65, 3.88066)
        (70, 4.0572)
        (75, 4.25593)
        (80, 4.46218)
        (90, 4.83623)
        (95, 4.96028)
        (99, 4.99923)
 };
    \addlegendentry{\small{$\psi=45^o$}}

    \addplot[
	color=black,
    only marks,
    mark=o]
	coordinates {
        (1, 4.99802)
        (5, 4.90505)
        (10, 4.62453)
        (20, 3.81737)
        (30, 2.99858)
        (40, 2.42575)
        (45, 2.27918)
        (50, 2.23425)
        (55, 2.27918)
        (60, 2.42575)
        (70, 2.99858)
        (80, 3.81737)
        (90, 4.62453)
        (95, 4.90505)
        (99, 4.99802)
 };
    \addlegendentry{\small{$\psi=60^o$}}

    \addplot[
	color=black,
    only marks,
    mark=star]
	coordinates {
        (1, 4.98838)
        (5, 4.56648)
        (10, 3.54352)
        (20, 1.51206)
        (30, 0.483065)
        (40, 0.180838)
        (45, 0.136164)
        (50, 0.123455)
        (55, 0.136164)
        (60,  0.180838)
        (70,  0.483065)
        (80, 1.51206)
        (90, 3.54352)
        (95, 4.56648)
        (99, 4.98838)
    };
    \addlegendentry{\small{$\psi=75^o$}}

    \addplot[
	color=black]
	coordinates {
        (1, 4.99802)
        (5, 4.90505)
        (10, 4.62453)
        (20, 3.81737)
        (30, 2.99858)
        (40, 2.42575)
        (45, 2.27918)
        (50, 2.23425)
        (55, 2.27918)
        (60, 2.42575)
        (70, 2.99858)
        (80, 3.81737)
        (90, 4.62453)
        (95, 4.90505)
        (99, 4.99802)
 };

  \addplot[
	color=black]
	coordinates {
        (1, 4.98838)
        (2, 4.93826)
        (4, 4.72152)
        (5, 4.56648)
        (7, 4.19277)
        (10, 3.54352)
        (15, 2.43864)
        (17, 2.03779)
        (20, 1.51206)
        (22, 1.21879)
        (25, 0.866338)
        (27, 0.68553)
        (30, 0.483065)
        (35, 0.280428)
        (40, 0.180838)
        (45, 0.136164)
        (50, 0.123455)
        (55, 0.136164)
        (60,  0.180838)
        (65, 0.280428)
        (70,  0.483065)
        (73, 0.68553)
        (75, 0.866338)
        (78, 1.21879)
        (80, 1.51206)
        (85, 2.43864)
        (90, 3.54352)
        (93, 4.19277)
        (95, 4.56648)
        (96, 4.72152)
        (98, 4.93826)
        (99, 4.98838)
    };

    \addplot[
	color=red]
	coordinates {
    (1, 4.99707)
    (3, 4.95685)
    (5, 4.86758)
    (10, 4.48857)
    (12, 4.29675)
    (15, 3.98558)
    (17, 3.76933)
    (20, 3.44166)
    (22, 3.22601)
    (25, 2.91462)
    (27, 2.71912)
    (30, 2.44999)
    (32, 2.30055)
    (35, 2.07978)
    (40, 1.81785)
    (42, 1.74343)
    (45, 1.66403)
    (47, 1.63577)
    (50, 1.61346)
    (53, 1.63577)
    (55, 1.66403)
    (58, 1.74343)
    (60, 1.81785)
    (65, 2.07978)
    (68, 2.30055)
    (70, 2.44999)
    (73, 2.71912)
    (75, 2.91462)
    (78, 3.22601)
    (80, 3.44166)
    (83, 3.76933)
    (85, 3.98558)
    (88, 4.29675)
    (90, 4.48857)
    (95, 4.86758)
    (97, 4.95685)
    (99, 4.99707)
    };
    
    \addplot[
    only marks,
	color=red,
    mark=square]
	coordinates {
    (1, 4.99707)
    (3, 4.95685)
    (5, 4.86758)
    (10, 4.48857)
    (12, 4.29675)
    (15, 3.98558)
    (17, 3.76933)
    (20, 3.44166)
    (22, 3.22601)
    (25, 2.91462)
    (27, 2.71912)
    (30, 2.44999)
    (32, 2.30055)
    (35, 2.07978)
    (40, 1.81785)
    (42, 1.74343)
    (45, 1.66403)
    (47, 1.63577)
    (50, 1.61346)
    (53, 1.63577)
    (55, 1.66403)
    (58, 1.74343)
    (60, 1.81785)
    (65, 2.07978)
    (68, 2.30055)
    (70, 2.44999)
    (73, 2.71912)
    (75, 2.91462)
    (78, 3.22601)
    (80, 3.44166)
    (83, 3.76933)
    (85, 3.98558)
    (88, 4.29675)
    (90, 4.48857)
    (95, 4.86758)
    (97, 4.95685)
    (99, 4.99707)
    };

    \addplot[
    only marks,
	color=red,
    mark=o]
	coordinates {
    (1, 4.99274)
    (3, 4.89902)
    (5, 4.70423)
    (7, 4.43575)
    (10, 3.94825)
    (12, 3.59385)
    (15, 3.05271)
    (17, 2.70077)
    (20, 2.20636)
    (22, 1.90798)
    (25, 1.51668)
    (27, 1.29588)
    (30, 1.02369)
    (32, 0.878739)
    (35, 0.708802)
    (40, 0.526121)
    (45, 0.433588)
    (47, 0.415458)
    (50, 0.405512)
    (53, 0.415458)
    (55, 0.433588)
    (60, 0.526121)
    (65, 0.708802)
    (68, 0.878739)
    (70, 1.02369)
    (73, 1.29588)
    (75, 1.51668)
    (78, 1.90798)
    (80, 2.20636)
    (83, 2.70077)
    (85, 3.05271)
    (88, 3.59385)
    (90, 3.94825)
    (93, 4.43575)
    (95, 4.70423)
    (97, 4.89902)
    (99, 4.99274)
    };

    \addplot[
	color=red]
	coordinates {
    (1, 4.99274)
    (3, 4.89902)
    (5, 4.70423)
    (7, 4.43575)
    (10, 3.94825)
    (12, 3.59385)
    (15, 3.05271)
    (17, 2.70077)
    (20, 2.20636)
    (22, 1.90798)
    (25, 1.51668)
    (27, 1.29588)
    (30, 1.02369)
    (32, 0.878739)
    (35, 0.708802)
    (40, 0.526121)
    (45, 0.433588)
    (47, 0.415458)
    (50, 0.405512)
    (53, 0.415458)
    (55, 0.433588)
    (60, 0.526121)
    (65, 0.708802)
    (68, 0.878739)
    (70, 1.02369)
    (73, 1.29588)
    (75, 1.51668)
    (78, 1.90798)
    (80, 2.20636)
    (83, 2.70077)
    (85, 3.05271)
    (88, 3.59385)
    (90, 3.94825)
    (93, 4.43575)
    (95, 4.70423)
    (97, 4.89902)
    (99, 4.99274)
    };

    \addplot[
    only marks,
	color=red,
    mark=star]
	coordinates {
        (1, 4.96034)
        (2, 4.80742)
        (3, 4.55545)
        (4, 4.2333)
        (5, 3.8587)
        (6, 3.47769)
        (7, 3.08213)
        (8, 2.69323)
        (9, 2.32078)
        (10, 1.97204)
        (12, 1.39188)
        (15, 0.70031)
        (17, 0.415413)
        (18, 0.31265)
        (19, 0.231831)
        (20, 0.169536)
        (22, 0.0872849)
        (25, 0.0299153)
        (27, 0.0141862)
        (30, 0.0045645)
        (32, 0.00216392)
        (35, 0.000742266)
        (37, 0.000384154)
        (40, 0.000160386)
        (42, 0.0000988394)
        (45, 0.0000570545)
        (47, 0.0000452283)
        (50, 0.0000395573)
        (53, 0.0000452283)
        (55, 0.0000570545)
        (58, 0.0000988394)
        (60, 0.000160386)
        (63, 0.000384154)
        (65, 0.000742266)
        (68, 0.00216392)
        (70, 0.0045645)
        (73, 0.0141862)
        (75, 0.0299153)
        (78, 0.0872849)
        (80, 0.169536)
        (81, 0.231831)
        (82, 0.31265)
        (83, 0.415413)
        (85, 0.70031)
        (88, 1.39188)
        (90, 1.97204)
        (91, 2.32078)
        (92, 2.69323)
        (93, 3.08213)
        (94, 3.47769)
        (95, 3.8587)
        (96, 4.2333)
        (97, 4.55545)
        (98, 4.80742)
        (99, 4.96034)
    };
    \addplot[
	color=red]
	coordinates {
        (1, 4.96034)
        (2, 4.80742)
        (3, 4.55545)
        (4, 4.2333)
        (5, 3.8587)
        (6, 3.47769)
        (7, 3.08213)
        (8, 2.69323)
        (9, 2.32078)
        (10, 1.97204)
        (12, 1.39188)
        (15, 0.70031)
        (17, 0.415413)
        (18, 0.31265)
        (19, 0.231831)
        (20, 0.169536)
        (22, 0.0872849)
        (25, 0.0299153)
        (27, 0.0141862)
        (30, 0.0045645)
        (32, 0.00216392)
        (35, 0.000742266)
        (37, 0.000384154)
        (40, 0.000160386)
        (42, 0.0000988394)
        (45, 0.0000570545)
        (47, 0.0000452283)
        (50, 0.0000395573)
        (53, 0.0000452283)
        (55, 0.0000570545)
        (58, 0.0000988394)
        (60, 0.000160386)
        (63, 0.000384154)
        (65, 0.000742266)
        (68, 0.00216392)
        (70, 0.0045645)
        (73, 0.0141862)
        (75, 0.0299153)
        (78, 0.0872849)
        (80, 0.169536)
        (81, 0.231831)
        (82, 0.31265)
        (83, 0.415413)
        (85, 0.70031)
        (88, 1.39188)
        (90, 1.97204)
        (91, 2.32078)
        (92, 2.69323)
        (93, 3.08213)
        (94, 3.47769)
        (95, 3.8587)
        (96, 4.2333)
        (97, 4.55545)
        (98, 4.80742)
        (99, 4.96034)
    };
 \end{axis}
\end{tikzpicture}}
	\vspace{-0.25cm}
	\caption{$D/W$ vs $d_1$ for different values of $\psi$ assuming $\kappa_t=\kappa_r=0$ (black), $\kappa_t=\kappa_r=0.1$ (red) and $d_1+d_2=100\,\rm{m}$.}
	\label{Fig:D_vs_d}
\end{figure}
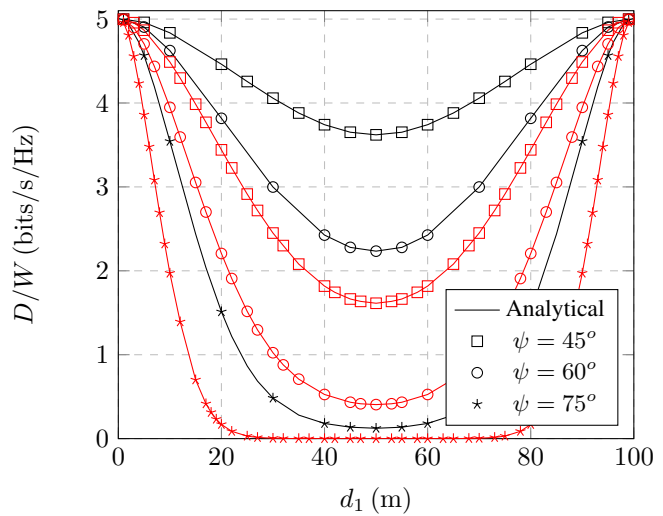

Fig.~\ref{Fig:D_vs_d} depicts $D/W$ as a function of $d_1$ for different values of $\psi$ and levels of hardware imperfections, assuming moderate fog conditions, $d_1+d_2=100\,\rm{m}$, $\frac{P_s}{\sigma_n^2}=80\,\rm{dB}$, $f=100\,\rm{GHz}$, $r_t=5\,\rm{bits/s/Hz}$, $G_{t}=G_r=50\,\rm{dBi}$, $\psi=\pi/4$, $T=20^{o}\rm{C}$, $P=101300.0\,\rm{Pa}$, and  $l_h=l_v=1\,\rm{m}$. The water vapor is set to $7.5\,\rm{g/m^3}$. From this figure, we observe that for given $\psi$, $\kappa_t=\kappa_r$ and $d_{1}<\frac{d_1+d_2}{2}$, as $d_1$ increases, the outage probability also increases; thus, $D/W$ decreases. On the other hand, for fixed $\psi$, $\kappa_t=\kappa_r$ and $d_{1}>\frac{d_1+d_2}{2}$, as $d_1$ increases, the outage probability decreases; hence, $D/W$ increases. For $d_1=\frac{d_1+d_2}{2}$, $D/W$ is minimized. Notice that this observation contradicts Fig.~\ref{Fig:OP_vs_d1}. This is because in Fig.~\ref{Fig:OP_vs_d1}, the impact of geometric losses was neglected. This indicates the importance of taking into account both the deterministic and stochastic phenomena  that affect the performance of the HRIS-empowered wireless THz system. Moreover, it becomes evident that the maximum $D/W$ is achieved for $d_1\to 0$ and $d_1\to d_1+d_2$. In other words, the optimal placement of the HRIS is as near either the TX or the RX as possible. Of note, this observation is in line with the results of~\cite{9386246}. Likewise, we observe that for fixed $d_1$ and  $\kappa_t=\kappa_r$, as $\psi$ increases, $D/W$ decreases. For example, for $d_1=50\,\rm{m}$ and $\kappa_t=\kappa_r=0.1$, as $\psi$ increases from $45^{o}$ to $75^{o}$, $D/W$ decreases about two orders of magnitude. Finally, the deterministic impact of hardware imperfections is revealed by this figure. In more detail, for given $d_1$ and $\psi$, as the level of transceiver hardware imperfections increases, $D/W$ decreases.      

\section{Conclusion}

In this paper, we have investigated the outage and throughput performance of HRIS-empowered THz wireless systems that suffer from the joint impact of transceiver hardware imperfections and fog. In more detail, after providing the methodology to evaluate the end-to-end geometric losses and characterizing the stochastic nature of the channel in terms of PDF and CDF, we have extracted novel closed-form expressions for the system's outage probability and throughput. These expressions have revealed a number of engineering insights, such as the existence of transmission windows, the optimum placement of the HRIS, the need to design transceivers with low noise figure, as well as the severe impact of transceiver hardware imperfections and fog. Therefore, they are expected to play a key role in the design of such systems.  

\section*{Appendices}
\section*{Appendix A}
\section*{Proof of Theorem 1}
Since $h_1$ and $h_2$ are independent random variables (RVs), the PDF of $A$ can be expressed~as
\begin{align}
    f_A(x) = \int_{x}^{1} \frac{1}{y} f_{h_1}(y)\,f_{h_2}\left(\frac{x}{y}\right)\,\rm{d}y. 
    \label{Eq:f_A_def}
\end{align}
By applying~\eqref{Eq:f_h_i} into~\eqref{Eq:f_A_def}, we obtain
\begin{align}
    f_A(x) = \frac{\zeta_1^{k_1}\,\zeta_2^{k_2}}{\Gamma(k_1)\,\Gamma(k_2)} x^{\zeta_2-1}\,&\int_{x}^{1} y^{\zeta_1-\zeta_2-1} \left(\ln\left(\frac{1}{y}\right)\right)^{k_1-1}  
    \nonumber \\ & \times
    \left(\ln\left(\frac{y}{x}\right)\right)^{k_2-1}\,\rm{d}y,
\end{align}
or equivalently
\begin{align}
    f_A(x) = \frac{\zeta_1^{k_1}\,\zeta_2^{k_2}}{\Gamma(k_1)\,\Gamma(k_2)} &x^{\zeta_2-1}\,\int_{x}^{1} y^{\zeta_1-\zeta_2-1} \left(-\ln\left({y}\right)\right)^{k_1-1}  
    \nonumber \\ & \times
    \left(\ln\left({y}\right)-\ln\left(x\right)\right)^{k_2-1}\,\rm{d}y.
    \label{Eq:f_A_step_2}
\end{align}
By setting $\ln(y) = t$,~\eqref{Eq:f_A_step_2} can be rewritten~as
\begin{align}
    f_A(x) = \frac{\zeta_1^{k_1}\,\zeta_2^{k_2}}{\Gamma(k_1)\,\Gamma(k_2)} x^{\zeta_2-1}\,&\int_{\ln(x)}^{0}  (-t)^{k_1-1}\,\left(\exp(t)\right)^{\zeta_1-\zeta_2}\nonumber \\ & \times\left(t-\ln(x)\right)^{k_2-1}\,\rm{d}t,
\end{align}
which, by applying~\cite[eq. (3.383/1)]{B:Gra_Ryz_Book}, yields~\eqref{Eq:f_A}.

The CDF of $A$ can be evaluated~as
\begin{align}
    F_{A}(x)=\int_{0}^{x} f_{A}(x)\,\rm{d}x,
\end{align}
which, by applying~\eqref{Eq:f_A}, can be rewritten~as
\begin{align}
    F_{A}(x)= (-1)^{k_1+k_2=1}\frac{\zeta_1^{k_1}\,\zeta_2^{k_2}}{\Gamma\left(k_1\right) \Gamma\left(k_2\right)}\,\mathcal{J},
    \label{Eq:F_A_s1}
\end{align}
where
\begin{align}
    \mathcal{J} = \int_{0}^{x} & y^{\zeta_2-1} \left(\ln(y)\right)^{k_1+k_2-1}
    \nonumber \\ & \times
    \,_1F_1\left(k_1;k_1+k_2;\left(\zeta_1-\zeta_2\right)\,\ln(y)\right)\,\rm{d}y.
    \label{Eq:J}
\end{align}
By setting $\ln(y)=t$,~\eqref{Eq:J} can be rewritten~as
\begin{align}
    \mathcal{J} = - \int_{\ln(y)}^{\infty}& t^{k_1+k_2-1}\,\exp\left(\zeta_2\, t\right)\nonumber \\ & \times
    \,_1F_1\left(k_1;k_1+k_2;\left(\zeta_1-\zeta_2\right)\,t\right)\,\rm{d}t.
    \label{Eq:J_s2}
\end{align}
With the aid of~\cite[eq. (9.14/1)]{B:Gra_Ryz_Book},~\eqref{Eq:J_s2} can be equivalently expressed~as
\begin{align}
\mathcal{J} = - \sum_{n=0}^{\infty} \frac{(k_1)_n\left(\zeta_1-\zeta_2\right)^{n}}{(k_1+k_2)_n\, n!} \mathcal{K},
\label{Eq:J_s3}
\end{align}
where 
\begin{align}
    \mathcal{K} = \int_{\ln(x)}^{\infty} t^{k_1+k_2+n-1}\,\exp\left(\zeta_2\,t\right)\,\rm{d}t.
    \label{Eq:K}
\end{align}
Next, by employing~\cite[eq. (1.211/1)]{B:Gra_Ryz_Book},~\eqref{Eq:K} can be written~as
\begin{align}
    \mathcal{K}=\sum_{m=0}^{\infty} \frac{\zeta_2^m}{m!}\,\int_{\ln(y)}^{\infty} t^{k_1+k_2+n+m-1}\,\rm{d}t,
\end{align}
which, by applying~\cite[eq. (3.191/1]{B:Gra_Ryz_Book}, can be rewritten~as
\begin{align}
    \mathcal{K}=-\sum_{m=0}^{\infty} \frac{\zeta_2^m}{m!}\,\frac{1}{k_1+k_2+m+n}\,\left(\ln(x)\right)^{k_1+k_2+m+n}.
    \label{Eq:K_s2}
\end{align}

By applying~\eqref{Eq:K_s2} into~\eqref{Eq:J_s3}, we obtain
\begin{align}
    \mathcal{J}=\left(\ln\left(x\right)\right)^{k_1+k_2}\,\mathcal{L},
    \label{Eq:J_s4}
\end{align}
where 
\begin{align}
    \mathcal{L}=\sum_{n=0}^{\infty}\sum_{m=0}^{\infty} & \frac{1}{k_1+k_2+m+n}\,\frac{(k_1)_n}{(k_1+k_2)_n} \frac{1}{n!}\frac{1}{m!}
    \nonumber \\ & \times
\left(\left(\zeta_1-\zeta_2\right)\ln(x)\right)^{n}\,\left(\zeta_2\ln(x)\right)^{m}.
    \label{Eq:L}
\end{align}
However, 
\begin{align}
    \frac{1}{k_1+k_2+m+n} = \frac{\left(k_1+k_2+m+n-1\right)!}{\left(k_1+k_2+m+n\right)!},
\end{align}
or equivalently
\begin{align}
    \frac{1}{k_1+k_2+m+n} = \frac{\Gamma\left(k_1+k_2+m+n-1\right)}{\Gamma\left(k_1+k_2+m+n\right)},
\end{align}
or
\begin{align}
    \frac{1}{k_1+k_2+m+n} = \frac{(k_1+k_2)_{m+n}\Gamma(m+n)}{(k_1+k_2+1)_{m+n}\Gamma(m+n+1)},
\end{align}
which can be rewritten~as
\begin{align}
    \frac{1}{k_1+k_2+m+n} = \frac{(k_1+k_2)_{m+n}}{(k_1+k_2+1)_{m+n}(k_1+k_2)_1)}.
    \label{Eq:frac_1_k_1_k_2}
\end{align}
Moreover, by accounting for the fact that $(k_1+k_2)_1=k_1+k_2$,~\eqref{Eq:frac_1_k_1_k_2} can be written~as 
\begin{align}
    \frac{1}{k_1+k_2+m+n} = \frac{1}{k_1+k_2}\frac{(k_1+k_2)_{m+n}}{(k_1+k_2+1)_{m+n}}.
    \label{Eq:frac_1_k_1_k_2_s2}
\end{align}
By applying~\eqref{Eq:frac_1_k_1_k_2_s2} to~\eqref{Eq:L}, we get
\begin{align}
    \mathcal{L}=\frac{1}{k_1+k_2} \sum_{n=0}^{\infty}&\sum_{m=0}^{\infty} \frac{(k_1+k_2)_{m+n}}{(k_1+k_2+1)_{m+n}} \,\frac{(k_1)_n}{(k_1+k_2)_n} \frac{1}{n!}\frac{1}{m!}
    \nonumber \\ & \times
\left(\left(\zeta_1-\zeta_2\right)\ln(x)\right)^{n}\,\left(\zeta_2\ln(x)\right)^{m}.
\end{align}
or equivalently
\begin{align}
    \mathcal{L} = \frac{\rm{H}_{A}\left(k_1+k_2, k_1; k_1+k_2+1; \left(\zeta_1-\zeta_2\right)\ln(x),\zeta_2\ln(x)\right)}{k_1+k_2}.
    \label{Eq:L_final}
\end{align}
By applying~\eqref{Eq:L_final} to~\eqref{Eq:J_s4}, we obtain
\begin{align}
    &\mathcal{J}=\frac{\left(\ln\left(x\right)\right)^{k_1+k_2}}{k_1+k_2}
    \nonumber \\ & \times
    \rm{H}_{A}\left(k_1+k_2, k_1; k_1+k_2+1; \left(\zeta_1-\zeta_2\right)\ln(x),\zeta_2\ln(x)\right),
    \label{Eq:J_s5}
\end{align}
Finally, by employing~\eqref{Eq:J_s5} to~\eqref{Eq:F_A_s1}, we obtain~\eqref{Eq:F_A}. This concludes the proof. 

\section*{Appendix B}
\section*{Proof of Proposition 1}
By applying~\eqref{Eq:gamma} to~\eqref{Eq:Po_def}, we obtain 
\begin{align}
    P_o\left(\gamma_{\rm{th}}\right) = \Pr\left(\frac{A^2}{A^2\left(\kappa_t^2+\kappa_r^2\right)+\frac{1}{\rho}}\leq \gamma_{\rm{th}}\right),
\end{align}
or equivalently
\begin{align}
    P_o\left(\gamma_{\rm{th}}\right) = \Pr\left(A^{2}\left(1-\gamma_{\rm{th}}\left(\kappa_t^2+\kappa_r^2\right)\right)\leq \frac{\gamma_{\rm{th}}}{\rho}\right).
    \label{Eq:P_o_s2}
\end{align}
By assuming that $1-\gamma_{\rm{th}}\left(\kappa_t^2+\kappa_r^2\right)>0$ or equivalently
\begin{align}
    \gamma_{\rm{th}} < \frac{1}{\kappa_t^{2}+\kappa_r^2},
    \label{Eq:gamma_th_in_1}
\end{align}
\eqref{Eq:P_o_s2} can be rewritten~as
\begin{align}
    P_o\left(\gamma_{\rm{th}}\right) = \Pr\left(A \leq \sqrt{\frac{\gamma_{\rm{th}}}{\rho}}\,\frac{1}{\sqrt{1-\gamma_{\rm{th}}\left(\kappa_t^2+\kappa_r^2\right)}}\right),
\end{align}
or
\begin{align}
    P_o\left(\gamma_{\rm{th}}\right) = F_{A}\left(\sqrt{\frac{\gamma_{\rm{th}}}{\rho}}\,\frac{1}{\sqrt{1-\gamma_{\rm{th}}\left(\kappa_t^2+\kappa_r^2\right)}}\right),
\end{align}
which, by applying~\eqref{Eq:F_A}, returns~\eqref{Eq:P_o_final}.

From~\eqref{Eq:F_A}, the following inequality should hold:
\begin{align}
   0 < \sqrt{\frac{\gamma_{\rm{th}}}{\rho}}\,\frac{1}{\sqrt{1-\gamma_{\rm{th}}\left(\kappa_t^2+\kappa_r^2\right)}} \leq 1
\end{align}
or equivalently
\begin{align}
   0 < {\frac{\gamma_{\rm{th}}}{\rho}}\,\frac{1}{{1-\gamma_{\rm{th}}\left(\kappa_t^2+\kappa_r^2\right)}} \leq 1
\end{align}
From~\eqref{Eq:gamma_th_in_1},  ${\frac{\gamma_{\rm{th}}}{\rho}}\,\frac{1}{{1-\gamma_{\rm{th}}\left(\kappa_t^2+\kappa_r^2\right)}}>0$ is always true. Likewise, in order for  ${\frac{\gamma_{\rm{th}}}{\rho}}\,\frac{1}{{1-\gamma_{\rm{th}}\left(\kappa_t^2+\kappa_r^2\right)}} \leq 1$ to be true, the following condition should be satisfied:
\begin{align}
    \gamma_{\rm{th}}\leq \frac{\rho}{\left(\kappa_t^2+\kappa_r^2\right)\rho+1}.
    \label{Eq:gamma_th_in_2}
\end{align}
By combining~\eqref{Eq:gamma_th_in_1} and~\eqref{Eq:gamma_th_in_2}, we obtain~\eqref{Eq:gamma_th_inequality}. This concludes the proof.
\balance
\bibliographystyle{IEEEtran}
\bibliography{IEEEabrv,References}

\begin{thebibliography}{10}
\providecommand{\url}[1]{#1}
\csname url@samestyle\endcsname
\providecommand{\newblock}{\relax}
\providecommand{\bibinfo}[2]{#2}
\providecommand{\BIBentrySTDinterwordspacing}{\spaceskip=0pt\relax}
\providecommand{\BIBentryALTinterwordstretchfactor}{4}
\providecommand{\BIBentryALTinterwordspacing}{\spaceskip=\fontdimen2\font plus
\BIBentryALTinterwordstretchfactor\fontdimen3\font minus
  \fontdimen4\font\relax}
\providecommand{\BIBforeignlanguage}[2]{{%
\expandafter\ifx\csname l@#1\endcsname\relax
\typeout{** WARNING: IEEEtran.bst: No hyphenation pattern has been}%
\typeout{** loaded for the language `#1'. Using the pattern for}%
\typeout{** the default language instead.}%
\else
\language=\csname l@#1\endcsname
\fi
#2}}
\providecommand{\BIBdecl}{\relax}
\BIBdecl

\bibitem{Boulogeorgos2018}
A.-A.~A. Boulogeorgos, A.~Alexiou, T.~Merkle, C.~Schubert, R.~Elschner,
  A.~Katsiotis, P.~Stavrianos, D.~Kritharidis, P.-K. Chartsias, J.~Kokkoniemi,
  M.~Juntti, J.~Lehtomaki, A.~Teixeira, and F.~Rodrigues, ``Terahertz
  technologies to deliver optical network quality of experience in wireless
  systems beyond 5{G},'' \emph{IEEE Commun. Mag.}, vol.~56, no.~6, pp.
  144--151, 2018.

\bibitem{WP:Wireless_Thz_system_architecture_for_networks_beyond_5G}
A.-A.~A. Boulogeorgos and {et al.}, ``Wireless terahertz system architectures
  for networks beyond {5G},'' TERRANOVA CONSORTIUM, White paper 1.0, Jul. 2018.

\bibitem{Boulogeorgos2021a}
A.-A.~A. Boulogeorgos, J.~M. Jornet, and A.~Alexiou, ``Directional terahertz
  communication systems for {6G: F}act check: {A} quantitative look,''
  \emph{{IEEE} Veh. Technol. Mag.}, pp. 2--10, Dec. 2021.

\bibitem{8782882}
K.~M.~S. Huq, S.~A. Busari, J.~Rodriguez, V.~Frascolla, W.~Bazzi, and D.~C.
  Sicker, ``Terahertz-enabled wireless system for beyond-5g ultra-fast
  networks: A brief survey,'' \emph{IEEE Network}, vol.~33, no.~4, pp. 89--95,
  2019.

\bibitem{8808165}
K.~K.O., W.~Choi, Q.~Zhong, N.~Sharma, Y.~Zhang, R.~Han, Z.~Ahmad, D.-Y. Kim,
  S.~Kshattry, I.~R. Medvedev, D.~J. Lary, H.-J. Nam, P.~Raskin, and I.~Kim,
  ``Opening terahertz for everyday applications,'' \emph{IEEE Communications
  Magazine}, vol.~57, no.~8, pp. 70--76, 2019.

\bibitem{8904434}
A.-A.~A. Boulogeorgos, E.~N. Papasotiriou, and A.~Alexiou, ``Analytical
  performance evaluation of thz wireless fiber extenders,'' in \emph{2019 IEEE
  30th Annual International Symposium on Personal, Indoor and Mobile Radio
  Communications (PIMRC)}, 2019, pp. 1--6.

\bibitem{5995306}
J.~M. Jornet and I.~F. Akyildiz, ``Channel modeling and capacity analysis for
  electromagnetic wireless nanonetworks in the terahertz band,'' \emph{IEEE
  Transactions on Wireless Communications}, vol.~10, no.~10, pp. 3211--3221,
  2011.

\bibitem{7086348}
K.~Yang, A.~Pellegrini, M.~O. Munoz, A.~Brizzi, A.~Alomainy, and Y.~Hao,
  ``Numerical analysis and characterization of thz propagation channel for
  body-centric nano-communications,'' \emph{IEEE Transactions on Terahertz
  Science and Technology}, vol.~5, no.~3, pp. 419--426, 2015.

\bibitem{7955066}
H.~Elayan, R.~M. Shubair, J.~M. Jornet, and P.~Johari, ``Terahertz channel
  model and link budget analysis for intrabody nanoscale communication,''
  \emph{IEEE Transactions on NanoBioscience}, vol.~16, no.~6, pp. 491--503,
  2017.

\bibitem{8568124}
J.~Kokkoniemi, J.~Lehtomäki, and M.~Juntti, ``Simplified molecular absorption
  loss model for 275–400 gigahertz frequency band,'' in \emph{12th European
  Conference on Antennas and Propagation (EuCAP 2018)}, 2018, pp. 1--5.

\bibitem{1638639}
M.~Uysal, J.~Li, and M.~Yu, ``Error rate performance analysis of coded
  free-space optical links over gamma-gamma atmospheric turbulence channels,''
  \emph{IEEE Transactions on Wireless Communications}, vol.~5, no.~6, pp.
  1229--1233, 2006.

\bibitem{Ma2015}
J.~Ma, F.~Vorrius, L.~Lamb, L.~Moeller, and J.~F. Federici, ``Experimental
  comparison of terahertz and infrared signaling in laboratory-controlled
  rain,'' \emph{Journal of Infrared, Millimeter, and Terahertz Waves}, vol.~36,
  no.~9, pp. 856--865, jun 2015.

\bibitem{9787400}
A.-A.~A. Boulogeorgos, J.~M. Riera, and A.~Alexiou, ``On the joint effect of
  rain and beam misalignment in terahertz wireless systems,'' \emph{IEEE
  Access}, vol.~10, pp. 58\,997--59\,012, 2022.

\bibitem{8751955}
L.~Cang, H.-K. Zhao, and G.-X. Zheng, ``The impact of atmospheric turbulence on
  terahertz communication,'' \emph{IEEE Access}, vol.~7, pp. 88\,685--88\,692,
  2019.

\bibitem{Taherkhani:20}
\BIBentryALTinterwordspacing
M.~Taherkhani, Z.~G. Kashani, and R.~Sadeghzadeh, ``Average bit error rate and
  channel capacity of terahertz wireless line-of-sight links with pointing
  errors under combined effects of turbulence and snow,'' \emph{Appl. Opt.},
  vol.~59, no.~33, pp. 10\,345--10\,356, Nov 2020. [Online]. Available:
  \url{https://opg.optica.org/ao/abstract.cfm?URI=ao-59-33-10345}
\BIBentrySTDinterwordspacing

\bibitem{Su:12}
\BIBentryALTinterwordspacing
K.~Su, L.~Moeller, R.~B. Barat, and J.~F. Federici, ``Experimental comparison
  of performance degradation from terahertz and infrared wireless links in
  fog,'' \emph{J. Opt. Soc. Am. A}, vol.~29, no.~2, pp. 179--184, Feb 2012.
  [Online]. Available:
  \url{https://opg.optica.org/josaa/abstract.cfm?URI=josaa-29-2-179}
\BIBentrySTDinterwordspacing

\bibitem{6971163}
Y.~Yang, M.~Mandehgar, and D.~R. Grischkowsky, ``Broadband thz signals
  propagate through dense fog,'' \emph{IEEE Photonics Technology Letters},
  vol.~27, no.~4, pp. 383--386, 2015.

\bibitem{Boulogeorgos2019}
A.-A.~A. Boulogeorgos, E.~N. Papasotiriou, and A.~Alexiou, ``Analytical
  performance assessment of {THz} wireless systems,'' \emph{IEEE Access},
  vol.~7, no.~1, pp. 1--18, Jan. 2019.

\bibitem{3gppR18}
\BIBentryALTinterwordspacing
{3GPP} (2022). {R}elease~18. [Online]. Available:
  \url{https://www.3gpp.org/release18}
\BIBentrySTDinterwordspacing

\bibitem{Mathaiou_COMMAG}
M.~Matthaiou, O.~Yurduseven, H.~Q. Ngo, D.~Morales-Jimenez, S.~L. Cotton, and
  V.~F. Fusco, ``The road to 6g: Ten physical layer challenges for
  communications engineers,'' \emph{IEEE Communications Magazine}, vol.~59,
  no.~1, pp. 64--69, 2021.

\bibitem{Renzo2020}
M.~Di~Renzo, A.~Zappone, M.~Debbah, M.-S. Alouini, C.~Yuen, J.~de~Rosny, and
  S.~Tretyakov, ``Smart radio environments empowered by reconfigurable
  intelligent surfaces: How it works, state of research, and the road ahead,''
  \emph{IEEE J. Sel. Areas Commun.}, vol.~38, no.~11, pp. 2450--2525, Nov.
  2020.

\bibitem{j:whitepaper2023}
``Reconfigurable intelligent surface technology,'' \emph{White Paper, RIS TECH
  Alliance}, Mar 2023.

\bibitem{9530717}
S.~Basharat, S.~A. Hassan, H.~Pervaiz, A.~Mahmood, Z.~Ding, and M.~Gidlund,
  ``Reconfigurable intelligent surfaces: Potentials, applications, and
  challenges for 6g wireless networks,'' \emph{IEEE Wireless Communications},
  vol.~28, no.~6, pp. 184--191, 2021.

\bibitem{9903378}
Z.~Chen, G.~Chen, J.~Tang, S.~Zhang, D.~K. So, O.~A. Dobre, K.-K. Wong, and
  J.~Chambers, ``Reconfigurable-intelligent-surface-assisted b5g/6g wireless
  communications: Challenges, solution, and future opportunities,'' \emph{IEEE
  Communications Magazine}, vol.~61, no.~1, pp. 16--22, 2023.

\bibitem{9919748}
N.~Chen, C.~Liu, H.~Jia, and M.~Okada, ``Intelligent reflecting surface aided
  network under interference toward 6g applications,'' \emph{IEEE Network},
  vol.~36, no.~4, pp. 18--27, 2022.

\bibitem{Basar2019}
E.~Basar, M.~Di~Renzo, J.~De~Rosny, M.~Debbah, M.-S. Alouini, and R.~Zhang,
  ``Wireless communications through reconfigurable intelligent surfaces,''
  \emph{IEEE Access}, vol.~7, pp. 116\,753--116\,773, Aug. 2019.

\bibitem{9433568}
F.~H. Danufane, M.~D. Renzo, J.~de~Rosny, and S.~Tretyakov, ``On the path-loss
  of reconfigurable intelligent surfaces: An approach based on green’s
  theorem applied to vector fields,'' \emph{IEEE Transactions on
  Communications}, vol.~69, no.~8, pp. 5573--5592, 2021.

\bibitem{9881509}
J.~Jeong, J.~H. Oh, S.~Y. Lee, Y.~Park, and S.-H. Wi, ``An improved path-loss
  model for reconfigurable-intelligent-surface-aided wireless communications
  and experimental validation,'' \emph{IEEE Access}, vol.~10, pp.
  98\,065--98\,078, 2022.

\bibitem{9837936}
W.~Tang, X.~Chen, M.~Z. Chen, J.~Y. Dai, Y.~Han, M.~D. Renzo, S.~Jin, Q.~Cheng,
  and T.~J. Cui, ``Path loss modeling and measurements for reconfigurable
  intelligent surfaces in the millimeter-wave frequency band,'' \emph{IEEE
  Transactions on Communications}, vol.~70, no.~9, pp. 6259--6276, 2022.

\bibitem{10159567}
K.~Rasilainen, T.~D. Phan, M.~Berg, A.~Pärssinen, and P.~J. Soh, ``Hardware
  aspects of sub-thz antennas and reconfigurable intelligent surfaces for 6g
  communications,'' \emph{IEEE Journal on Selected Areas in Communications},
  pp. 1--1, 2023.

\bibitem{9690474}
R.~Deng, B.~Di, H.~Zhang, D.~Niyato, Z.~Han, H.~V. Poor, and L.~Song,
  ``Reconfigurable holographic surfaces for future wireless communications,''
  \emph{IEEE Wireless Communications}, vol.~28, no.~6, pp. 126--131, 2021.

\bibitem{7109827}
G.~Oliveri, D.~H. Werner, and A.~Massa, ``Reconfigurable electromagnetics
  through metamaterials—a review,'' \emph{Proceedings of the IEEE}, vol. 103,
  no.~7, pp. 1034--1056, 2015.

\bibitem{nature}
S.~Venkatesh, X.~Lu, H.~Saeidi, and K.~Sengupta, ``A high-speed programmable
  and scalable terahertz holographic metasurface based on tiled cmos chips,''
  \emph{Nat. Electron.}, vol.~3, no.~12, pp. 785--793, 2020.

\bibitem{Nature_Light}
F.~Lan and \emph{et.al}, ``Real-time programmable metasurface for terahertz
  multifunctional wave front engineering,'' \emph{Light Sci. Appl.}, vol.~12,
  no.~8, pp. 1--12, Aug. 2023, paper No. 191.

\bibitem{Huang2020}
C.~Huang, S.~Hu, G.~C. Alexandropoulos, A.~Zappone, C.~Yuen, R.~Zhang, M.~D.
  Renzo, and M.~Debbah, ``Holographic {MIMO} surfaces for 6{G} wireless
  networks: Opportunities, challenges, and trends,'' \emph{IEEE Wireless
  Commun.}, vol.~27, no.~5, pp. 118--125, Jul. 2020.

\bibitem{Deng2021}
R.~Deng, B.~Di, H.~Zhang, Y.~Tan, and L.~Song, ``Reconfigurable holographic
  surface: Holographic beamforming for metasurface-aided wireless
  communications,'' \emph{IEEE Trans. Veh. Technol.}, vol.~70, no.~6, pp.
  6255--6259, Jun. 2021.

\bibitem{9374451}
Z.~Wan, Z.~Gao, F.~Gao, M.~D. Renzo, and M.-S. Alouini, ``Terahertz massive
  mimo with holographic reconfigurable intelligent surfaces,'' \emph{IEEE
  Transactions on Communications}, vol.~69, no.~7, pp. 4732--4750, 2021.

\bibitem{9903514}
J.~Dang, Z.~Zhang, Y.~Li, L.~Wu, B.~Zhu, and L.~Wang, ``Fast and arbitrary beam
  pattern design for ris-assisted terahertz wireless communication,''
  \emph{IEEE Transactions on Vehicular Technology}, vol.~72, no.~2, pp.
  2620--2625, 2023.

\bibitem{10087279}
R.~Su, L.~Dai, and D.~W.~K. Ng, ``Wideband precoding for ris-aided thz
  communications,'' \emph{IEEE Transactions on Communications}, vol.~71, no.~6,
  pp. 3592--3604, 2023.

\bibitem{10001180}
A.~P. Chrysologou, A.-A.~A. Boulogeorgos, N.~D. Chatzidiamantis, and
  A.~Alexiou, ``Outage analysis of holographic surface assisted downlink
  terahertz {NOMA},'' in \emph{Proc. IEEE Global Commun. Conf. (GLOBECOM)},
  Dec. 2022, pp. 5249--5254.

\bibitem{10177184}
A.~P. Chrysologou, A.-A.~A. Boulogeorgos, and N.~D. Chatzidiamantis, ``When
  thz-noma meets holographic reconfigurable intelligent surfaces,'' \emph{IEEE
  Communications Letters}, pp. 1--1, 2023.

\bibitem{B:Gra_Ryz_Book}
I.~S. Gradshteyn and I.~M. Ryzhik, \emph{Table of Integrals, Series, and
  Products}, 6th~ed.\hskip 1em plus 0.5em minus 0.4em\relax New York: Academic,
  2000.

\bibitem{CARLSON1963452}
\BIBentryALTinterwordspacing
B.~Carlson, ``Lauricella's hypergeometric function fd,'' \emph{Journal of
  Mathematical Analysis and Applications}, vol.~7, no.~3, pp. 452--470, 1963.
  [Online]. Available:
  \url{https://www.sciencedirect.com/science/article/pii/0022247X63900672}
\BIBentrySTDinterwordspacing

\bibitem{8936989}
O.~\"Ozdogan, E.~Bj\"ornson, and E.~G. Larsson, ``Intelligent reflecting
  surfaces: Physics, propagation, and pathloss modeling,'' \emph{IEEE Wireless
  Communications Letters}, vol.~9, no.~5, pp. 581--585, 2020.

\bibitem{CLOUGH1989229}
\BIBentryALTinterwordspacing
S.~Clough, F.~Kneizys, and R.~Davies, ``Line shape and the water vapor
  continuum,'' \emph{Atmospheric Research}, vol.~23, no.~3, pp. 229--241, 1989.
  [Online]. Available:
  \url{https://www.sciencedirect.com/science/article/pii/0169809589900203}
\BIBentrySTDinterwordspacing

\bibitem{RevModPhys.17.227}
\BIBentryALTinterwordspacing
J.~H. Van~Vleck and V.~F. Weisskopf, ``On the shape of collision-broadened
  lines,'' \emph{Rev. Mod. Phys.}, vol.~17, pp. 227--236, Apr 1945. [Online].
  Available: \url{https://link.aps.org/doi/10.1103/RevModPhys.17.227}
\BIBentrySTDinterwordspacing

\bibitem{8580934}
E.~N. Papasotiriou, J.~Kokkoniemi, A.-A.~A. Boulogeorgos, J.~Lehtomäki,
  A.~Alexiou, and M.~Juntti, ``A new look to 275 to 400 ghz band: Channel model
  and performance evaluation,'' in \emph{2018 IEEE 29th Annual International
  Symposium on Personal, Indoor and Mobile Radio Communications (PIMRC)}, 2018,
  pp. 1--5.

\bibitem{8123513}
K.~Tsujimura, K.~Umebayashi, J.~Kokkoniemi, J.~Lehtomäki, and Y.~Suzuki, ``A
  causal channel model for the terahertz band,'' \emph{IEEE Transactions on
  Terahertz Science and Technology}, vol.~8, no.~1, pp. 52--62, 2018.

\bibitem{itu}
{Radiocommunication Sector of International Telecommunication Union},
  ``Recommendation {ITU-R} {P.840-6:} attenuation due to clouds and fog,''
  2013.

\bibitem{9714471}
O.~S. Badarneh, ``Performance analysis of terahertz communications in random
  fog conditions with misalignment,'' \emph{IEEE Wireless Communications
  Letters}, vol.~11, no.~5, pp. 962--966, 2022.

\bibitem{B:Schenk-book}
T.~Schenk, \emph{{RF} Imperfections in High-Rate Wireless Systems}.\hskip 1em
  plus 0.5em minus 0.4em\relax The Netherlands: Springer, 2008.

\bibitem{6630485}
E.~Bjornson, M.~Matthaiou, and M.~Debbah, ``A new look at dual-hop relaying:
  Performance limits with hardware impairments,'' \emph{IEEE Transactions on
  Communications}, vol.~61, no.~11, pp. 4512--4525, 2013.

\bibitem{PhD:Boulogeorgos}
A.-A.~A. Boulogeorgos, ``Interference mitigation techniques in modern wireless
  communication systems,'' Ph.D. dissertation, Aristotle University of
  Thessaloniki, Thessaloniki, Greece, Sep. 2016.

\bibitem{Koenig2013}
S.~Koenig, D.~Lopez-Diaz, J.~Antes, F.~Boes, R.~Henneberger, A.~Leuther,
  A.~Tessmann, R.~Schmogrow, D.~Hillerkuss, R.~Palmer, T.~Zwick, C.~Koos,
  W.~Freude, O.~Ambacher, J.~Leuthold, and I.~Kallfass, ``Wireless sub-{THz}
  communication system with high data rate,'' \emph{Nat. Photonics}, vol.~7,
  pp. 977 EP--, Oct. 2013.

\bibitem{Boulogeorgos2020b}
A.-A.~A. Boulogeorgos and A.~Alexiou, ``Performance analysis of reconfigurable
  intelligent surface-assisted wireless systems and comparison with relaying,''
  \emph{IEEE Access}, vol.~8, pp. 94\,463--94\,483, May 2020.

\bibitem{8885655}
------, ``Analytical performance evaluation of beamforming under transceivers
  hardware imperfections,'' in \emph{2019 IEEE Wireless Communications and
  Networking Conference (WCNC)}, 2019, pp. 1--7.

\bibitem{9386246}
K.~Ntontin, A.-A.~A. Boulogeorgos, D.~G. Selimis, F.~I. Lazarakis, A.~Alexiou,
  and S.~Chatzinotas, ``Reconfigurable intelligent surface optimal placement in
  millimeter-wave networks,'' \emph{IEEE Open Journal of the Communications
  Society}, vol.~2, pp. 704--718, 2021.

\end{thebibliography}

\end{document}